\documentclass[aps,prb,twocolumn,amsmath,superscriptaddress,amssymb]{revtex4}
\usepackage{graphicx}
\usepackage{enumitem}
\usepackage[colorlinks,bookmarks=false,citecolor=red,linkcolor=blue,urlcolor=blue]{hyperref}

\newcommand{\X}[1]{\ensuremath{S^x_{#1}}}
\newcommand{\Y}[1]{\ensuremath{S^y_{#1}}}
\newcommand{\Z}[1]{\ensuremath{S^z_{#1}}}

\allowdisplaybreaks

\begin{document}

\title{Order by singularity in Kitaev clusters}
\author{Sarvesh Srinivasan}
\email{f2015191@pilani.bits-pilani.ac.in}
\affiliation{The Institute of Mathematical Sciences, HBNI, C I T Campus, Chennai 600113, India}
\affiliation{Birla Institute of Technology and Science, Pilani 333031, India}
\author{Subhankar Khatua}
\email{subhankark@imsc.res.in}
\affiliation{The Institute of Mathematical Sciences, HBNI, C I T Campus, Chennai 600113, India}
\author{G. Baskaran}
\email{baskaran@imsc.res.in}
\affiliation{The Institute of Mathematical Sciences, HBNI, C I T Campus, Chennai 600113, India}
\affiliation{Indian Institute of Technology Madras, Chennai 600036, India}
\affiliation{Perimeter Institute for Theoretical Physics, Waterloo, Ontario N2L 2Y5, Canada}
\author{R. Ganesh}
\email{ganesh@imsc.res.in}
\affiliation{The Institute of Mathematical Sciences, HBNI, C I T Campus, Chennai 600113, India}
\date{\today}

\begin{abstract}
The Kitaev model is a beautiful example of frustrated interactions giving rise to deep and unexpected phenomena. In particular, its classical version has remarkable properties stemming from exponentially large ground state degeneracy. Here, we present a study of magnetic clusters with spin-$S$ moments coupled by Kitaev interactions. We focus on two cluster geometries -- the Kitaev square and the Kitaev tetrahedron -- that allow us to explicitly enumerate all classical ground states. In both cases, the classical ground state space (CGSS) is large and self-intersecting, with non-manifold character. 
The Kitaev square has a CGSS of four intersecting circles that can be embedded in four dimensions. The tetrahedron CGSS consists of eight spheres embedded in six dimensions. 
In the semi-classical large-$S$ limit, we argue for effective low energy descriptions in terms of a single particle moving on these non-manifold spaces. Remarkably, at low energies, the particle is tied down in bound states formed around singularities at self-intersection points. In the language of spins, the low energy physics is determined by a distinct set of states that lies well below other eigenstates. These correspond to `Cartesian' states, a special class of classical ground states that are constructed from dimer covers of the underlying lattice. They completely determine the low energy physics despite being a small subset of the classical ground state space. 
This provides an example of order by singularity, where state selection becomes stronger upon approaching the classical limit.
\end{abstract}
%\pacs{}
                                 
\maketitle
\section{Introduction}
Frustrated magnetism is fertile ground for several interesting phenomena. This is typically best understood in the $S \rightarrow \infty$ limit where frustration gives rise to a large classical ground state degeneracy. The effects of this degeneracy persist even as we move away from the classical limit. Its most significant consequence is to determine the nature of ordering, if at all long range magnetic order emerges in the system. 
This selection of order by fluctuations is captured by the `order by disorder' (ObD) paradigm\cite{Chalker2011, Villain1980, Shender1982, Henley1989}.  
In the case of quantum fluctuations, this is typically captured by small $\mathcal{O}(1/S)$ corrections. They break the classical degeneracy by their zero point energies to give rise to ordering. Likewise, weak thermal fluctuations at low temperatures, can break degeneracy by allowing for varying entropies. Both lead to long range order in a fluctuation-selected ground state. 

A new selection paradigm, order by singularity (ObS),  has recently been proposed by two of the current authors. We briefly recapitulate its gist here; details can be found in Ref.~\onlinecite{Khatua2019}. We start with a general principle that holds in the semi-classical large-$S$ limit : the low energy physics of a cluster of quantum spins maps to that of a single particle moving on the classical ground state space (CGSS). In particular, the low-lying energy states of a spin cluster have a one-to-one relation with those of the corresponding single particle problem. This mapping can be seen from the spin path integral formulation combined with a large-$S$ semiclassical approach. However, this path integral-based argument can be carried out only in systems where the CGSS is a smooth manifold. Nevertheless, the mapping is conjectured to hold for systems with non-manifold CGSS' as well. As proof of principle, it was shown to hold true for the XY quadrumer. Remarkably, this example brings out a distinctive localization phenomenon arising from self-intersection in the CGSS. 
In the single particle picture, these self-intersection points or singularities mimic impurities to create bound states.The particle is then tied down in bound states at low energies, preventing ergodic sampling of the CGSS. For the magnetic cluster, this manifests as a preference for certain classical ground states over others.

As a mechanism for state selection, ObS can be distinguished from ObD as follows.  
As we approach the classical $S\rightarrow \infty$ limit, state selection due to ObS becomes stronger. This is because the mapping between the spin system and the single particle problem becomes exact in this limit.
In contrast, selection due to ObD weakens with increasing $S$, eventually vanishing in the classical limit. As quantum fluctuations are $\mathcal{O}(1/S)$ corrections, their effects diminish with increasing $S$.
In this article, we provide two new examples of ObS in clusters with Kitaev-like couplings. The small size of the clusters allows us to explicitly map out their CGSS'. In both cases, we find interesting CGSS topology with self-intersections. By mapping the spin problem to a particle moving on the CGSS, we find localization within bound states, heralding ObS. As this selection behaviour determines the physics at large $S$, it sheds light on the semiclassical behaviour of Kitaev models.

The Kitaev model was proposed in 2006 as an artificial system that allows for an exact solution in terms of free fermions and $\mathbb{Z}_2$ gauge fields\cite{Kitaev2006}. It describes spin-$1/2$ moments on a honeycomb lattice with nearest-neighbour Ising-like $x-x$, $y-y$ and $z-z$ bonds. The model has received tremendous interest from the point of view of fundamental physics\cite{Kitaev2009,Nussinov2013,Perreault2016,Zhou2017,Rao2017,Hermanns2018} as well as from a materials angle\cite{Rau2016,Takagi2019}.  
Several extensions of the model have been proposed to different lattices, couplings, etc. A particularly interesting extension is realized by promoting the spins to the semiclassical limit with $S\gg 1/2$. This leads to several interesting phenomena: an exponentially large classical ground state space\cite{Baskaran2008}, local plaquette-centred conserved quantities\cite{Baskaran2008}, equivalence to a height model\cite{Chandra2010}, power law correlations in certain variables\cite{Chandra2010}, 
spin liquid behaviour even in the semiclassical limit\cite{Rousochatzakis2018}, etc. The seeds of some of these features appear in a simple and tractable form in the two clusters that we study in this article.

The remainder of this article is structured as follows. In Sec.~\ref{sec.classical}, we review what is known about the Kitaev model in the classical and semi-classical limits. In the process, we recapitulate the definition of a cartesian state -- a key notion in following sections.  
We next discuss a toy problem of a particle moving on two intersecting circles in Sec.~\ref{sec.twocircles}. This sets the stage for studying Kitaev clusters, highlighting the key role of bound states. In Sec.~\ref{sec.clusters}, we introduce the two clusters and their symmetries. We move on to the Kitaev square in Sec.~\ref{sec.square} where we construct the CGSS, interpret its features and discuss the quantum spectrum. We also discuss two independent tests for the nature of the low energy states. We discuss the tetrahedron on similar lines in Sec.~\ref{sec.tetrahedron}. We conclude with a summary and discussion.

\begin{figure*}
\includegraphics[width=2\columnwidth]{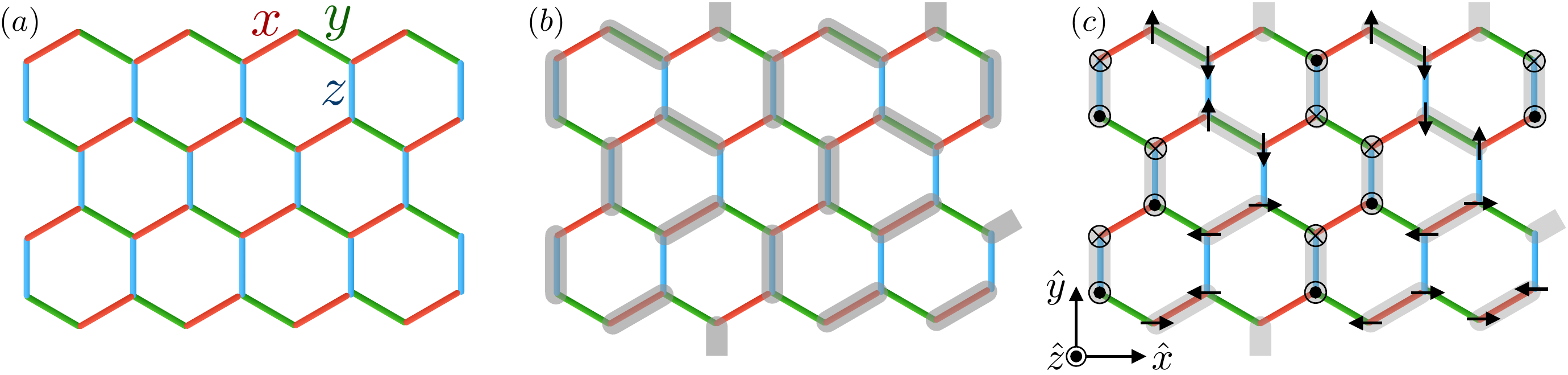}
\caption{(a) The Kitaev model on the honeycomb lattice. Bonds have $x-x$ (red), $y-y$ (green) or $z-z$ (blue) couplings depending on their orientation. 
(b) A dimer cover on the honeycomb lattice. (c) A cartesian state constructed from the dimer cover. On a dimer of type $u$ $~(u=x,y,z)$, the two spins are chosen to lie along $+u$ and $-u$ directions. There are two choices for each dimer corresponding to choosing one of the spins to point along $+u$.}
\label{fig.cartesian}
\end{figure*}

\section{The Kitaev model in the classical limit}
\label{sec.classical}

The Kitaev model is described by the Hamiltonian 
\begin{eqnarray}
H = K\Big[
\sum_{\langle ij \rangle_x} \hat{S}_{i}^x ~\hat{S}_{j}^x
+ \sum_{\langle ij \rangle_y} \hat{S}_{i}^y ~\hat{S}_{j}^y
+ \sum_{\langle ij \rangle_z} \hat{S}_{i}^z ~\hat{S}_{j}^z \Big],
\label{eq.Ham}
\end{eqnarray}
with $\langle ij \rangle_{x/y/z}$ representing nearest neighbour bonds on the honeycomb lattice. There are three types of bonds with Ising-like couplings in the $x$, $y$ and $z$ components respectively, as shown in Fig.~\ref{fig.cartesian}(left). We will assume $K>0$ for concreteness; the results and statements that follow can be easily modified to suit the $K<0$ case. 
While the original Kitaev model is formulated for $S=1/2$ moments, there is a growing body of work on this model with spins elevated to arbitrary $S$. In this section, we summarize the salient results that are known about the model in the classical $S\rightarrow \infty$ limit.

The seminal work of Baskaran, Sen and Shankar\cite{Baskaran2008} (BSS hereafter) brought out, among other things, rich structure in the CGSS of the Kitaev model.
A convenient starting point to understand this structure is the notion of `cartesian' states. To define a cartesian state, we begin with a nearest-neighbour dimer cover of the honeycomb lattice, as shown in Fig.~\ref{fig.cartesian}(centre). On each dimer, we take the two spins at its end points and align them as follows. One spin is aligned along the `bond direction' while the other is placed in the opposite direction. For example, on a dimer on an x-bond, one spin is taken to point along the $\hat{x}$ direction with the other pointing along $-\hat{x}$. This gives the lowest energy contribution from this bond. There are two such spin configurations on each dimer, leading to an exponentially large number of possibilities for a given dimer cover. An example spin configuration is shown in Fig.~\ref{fig.cartesian}(right) corresponding to the dimer cover shown in Fig.~\ref{fig.cartesian}(centre).
In addition to the degeneracy of spin alignments, we have an exponentially large number of choices for a dimer cover on the underlying honeycomb lattice. Clearly, the set of all cartesian states is very large, scaling exponentially with the system size. Remarkably, every cartesian state is a ground state of the classical Kitaev Hamiltonian. Even more remarkably, a given cartesian state can be smoothly transformed into other cartesian states via a continuous one-parameter transformation. All intermediate states are also classical ground states of the problem. With this picture, BSS envisages the CGSS as `an exponentially large number of isolated points connected by flat valleys'.

A more rigorous discussion of the CGSS was given by Chandra et. al. through a mapping to electrostatics\cite{Chandra2010}. This potentially reveals new classical ground states beyond those enumerated by BSS. However, it is difficult to explicitly construct these states and to determine their connectivity. Chandra et.  al. draws several conclusions: (a) the CGSS is an $(N+1)$-dimensional manifold, where $2N$ is the number of sites, (b) the cartesian states are `extrema' in the CGSS, (c) in the zero-temperature partition function, the cartesian states contribute a larger weight compared to other states, and (d) there is no selection of states by fluctuations in the $T\rightarrow 0$ limit. In this article, we present a detailed study of two clusters wherein these types of issues can be more readily examined.

More recently, Rousochatzakis et. al. provide an illuminating discussion of the Kitaev problem in the large $S$ limit\cite{Rousochatzakis2018}. They introduce a new parametrization for the classical ground state space. In the limit of large-$S$, weak quantum fluctuations play a dramatic role by `selecting' a subset of this space, constructed from star-like dimer covers on the honeycomb lattice. The low-energy physics is restricted to fluctuations within this sector. It takes a remarkable form, mapping to the toric code problem on the Kagome lattice. The $\mathbb{Z}_2$ gauge theory structure is inherited from local conserved quantities that were first pointed out by BSS. The current article, albeit restricted to small clusters, points out selection effects beyond the quantum fluctuation paradigm, arising from the topology of the ground state space itself.

Our study of Kitaev clusters can be seen as a progression of earlier work extending the Kitaev structure to systems beyond the honeycomb lattice. 
Kitaev physics has been studied in one-dimensional and even three-dimensional systems\cite{Saket2010,Mandal2009,Trebst2017}. The essential requirement is three-fold coordination of nearest neighbour bonds. Lattices with six-fold coordination, such as the triangular lattice, can also host Kitaev-type couplings\cite{Jackeli2015,Avella2018}.
Significant insights have been gleaned from analysing these problems in the classical and semi-classical limits. This has also revealed new physics beyond the original Kitaev formulation. A particularly elegant example is the crystallization of $\mathbb{Z}_2$ vortices on the triangular lattice\cite{Rousochatzakis2016,Seabrook2019}.  
\section{Particle on two intersecting circles: a toy problem }
\label{sec.twocircles}
\begin{figure*}
\includegraphics[width=7in]{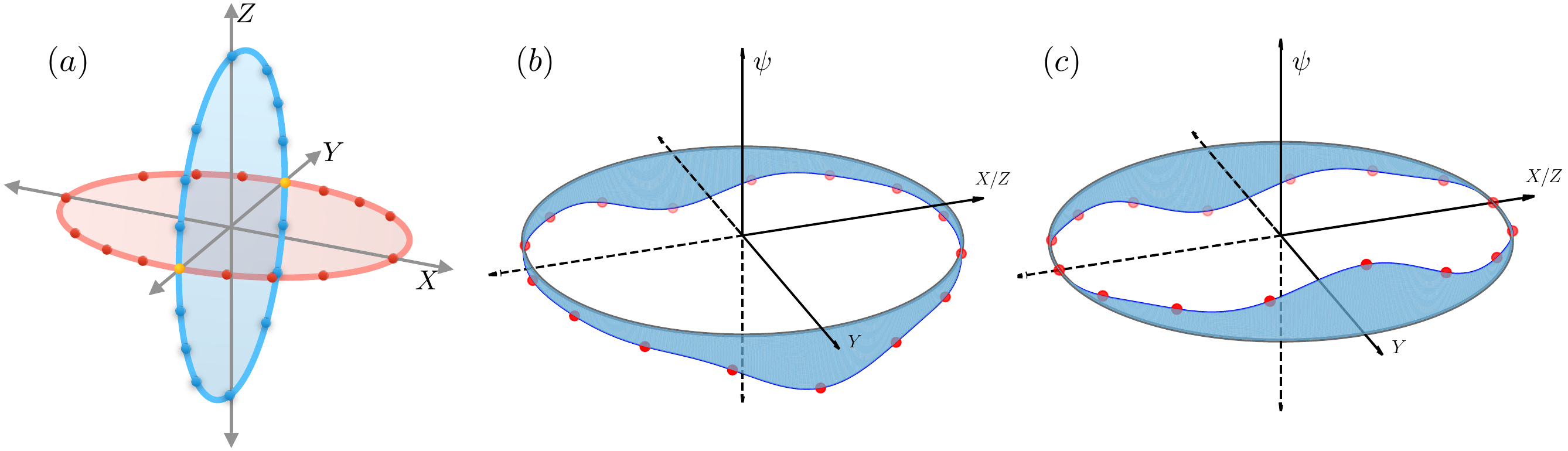}
\caption{(a) Space of two circles in orthogonal planes. The intersection points are shown in yellow. The space is discretized, with the particle allowed to hop between nearest neighbours. Note that a generic point has two neighbours, while the intersection points have four. (b) Numerically obtained ground state wavefunction with the discretization mesh chosen to have 16 sites on each circle. 
The base represents the XY and the YZ circles, as the wavefunction is the same on both circles. As the wavefunction is purely real, we represent it by the heights of red dots from the base. 
Note that wavefunction is peaked at the intersection points along the Y axis. (c) The numerically obtained first excited state. As with the ground state, this wavefunction is purely real and takes the same form on both circles.  
The wavefunction has opposite sign at the two intersection points ($\pm \hat{y}$).}
\label{fig.twocircles}
\end{figure*}

Before discussing the Kitaev problem at hand, we first discuss a simple example of dynamics on a non-manifold space. This sets the stage for discussions of Kitaev clusters in the following sections. We consider a space of two circles with unit radius that are centred at the origin, as shown in Fig.~\ref{fig.twocircles}(left). While the first circle lies in the XY plane, the second lies in the YZ plane. The circles intersect at two points, $\pm \hat{y}$. We refer to these as self-intersections as the space intersects itself at these points. This space is a `non-manifold': while it is one-dimensional at generic points, it does not have well-defined dimensionality in the vicinity of the self-intersection points. We now consider a single particle moving on this space. At generic points, the particle moves along one of the circles. At a self-intersection point, it is allowed to move from one circle to another. For reasons that are explained below, we are interested in the low energy behaviour of this particle, i.e., in stationary states with the lowest energy.

The eigenstates for this problem cannot be calculated using standard quantum mechanical tools, unlike, say, a particle on a single circle. For example, we cannot define a gradient operator on this space. We take an alternative approach by discretizing this space to build a tight binding Hamiltonian. As shown in Fig.~\ref{fig.twocircles}(a), a generic point is connected to two neighbours that lie on the same circle. In contrast, a self-intersection point is connected to four neighbouring sites, two on each circle. For a given mesh size (discretization), the spectrum can be easily obtained numerically. The resulting wavefunctions in the ground state and the first excited state are shown in Figs.~\ref{fig.twocircles}(b) and (c).

The two lowest energy states in the tight binding problem are qualitatively different from the other, higher energy, states. They are `bound states' that are localized around the self-intersection points. To see this, we consider the limit of dense discretization, where the self-intersection points are separated by a large number of intermediate points. Focussing on the vicinity of one self-intersection, we label sites as $(n,A/B)$, where $A$ and $B$ denote the two circles and the integer $n$ represents sites on each circle. We take $n=0$ to be the intersection point with $(0,A)\equiv (0,B)$. We propose an ansatz for the bound state given by 
\begin{eqnarray}
\psi_{n,A/B} =  \frac{1}{\mathcal{N}} \exp[-\alpha n],
\label{eq.bound}
\end{eqnarray}
where $\mathcal{N}$ is the normalization constant and $\alpha$ is a decay constant that is to be determined. This wavefunction is purely real. It is symmetrically distributed on the two circles, decaying exponentially as we move away from the self-intersection point.
Assuming that it is an eigenstate of the Hamiltonian with eigenvalue $E$, the Schr\"odinger equation at a generic site takes the form $E = -t (e^{\alpha} + e^{-\alpha})$. At the intersection point ,the Schr\"odinger equation takes the form $E = -4t e^{-\alpha}$. From these two relations, we find $\alpha =\frac{\ln3}{2}$.

To examine its bound nature, we compare it with unbound states in the problem. Away from the intersection point, the space resembles a circle. Eigenstates in this region resemble solutions on a circle with the dispersion relation, $E_{unbound} = -2 t \cos k$, where $k$ is the one-dimensional momentum quantum number. These states have energies in the range, $[-2t,2t]$. Crucially, the state in Eq.~\ref{eq.bound} lies below this window, with energy $ E_{bound} = -4t/\sqrt{3} \approx -2.3094 t$. This signifies that the bound state does not hybridize with delocalized modes. More importantly, it indicates that the bound state is the lowest energy state in the problem. 

In the full space with two circles, we have two bound states with one at each self-intersection point. When the discretization is not too dense, the bound state wavefunctions overlap in the intermediate region. This results in mixing which splits them into a symmetric and an anti-symmetric combination. The symmetric state, with lower energy, becomes the ground state, while the anti-symmetric state becomes the first excited state. Their wavefunctions are shown in Figs.~\ref{fig.twocircles} (b) and (c). The splitting between the symmetric and anti-symmetric state decreases as we make the discretization more dense. The individual bound state wavefunction in Eq.~\ref{eq.bound} has a localization length of $\frac{1}{\alpha} $ lattice spacings. If the number of intervening lattice points is increased, the overlap between the two bound states decreases. In the limit of very dense discretization, we have independent bound states that are sharply localized around the self-intersection points.

The tight binding results on the two-circle-space provide a framework to understand the results on spin clusters below. The spin clusters have CGSS' that are non-manifold spaces, analogous to the two-circle-space described here. In each cluster, the low energy physics maps to a particle moving on the corresponding CGSS. The spin quantum number, ${S}$, loosely corresponds to the denseness of the tight binding mesh. 
As $S$ increases, the mesh becomes denser with a larger number of sites. Such an interpretation for $S$ was given in Ref.~\onlinecite{Khatua2019} in the context of the XY quadrumer (see Tables I and II therein). As we will show in the sections below, the results in Kitaev clusters are also consistent with this interpretation. 
The two-circle problem brings out the following three aspects that carry over to the spin 
clusters: 
(i) The spectrum contains a set of low energy states that is well separated from other, higher energy, states. The number of such states is the same as the number of self-intersection points in the CGSS. (ii) These low energy states are admixtures of bound states that form around self-intersection points, e.g., the ground state is a symmetric combination of all bound states. Their wavefunctions are peaked at the intersection points. (iii) In the dense discretization limit, each low energy eigenstate is associated with one self-intersection point, being sharply localized in its vicinity.

\section{Kitaev clusters}
\label{sec.clusters}
\begin{figure}
\includegraphics[width=\columnwidth]{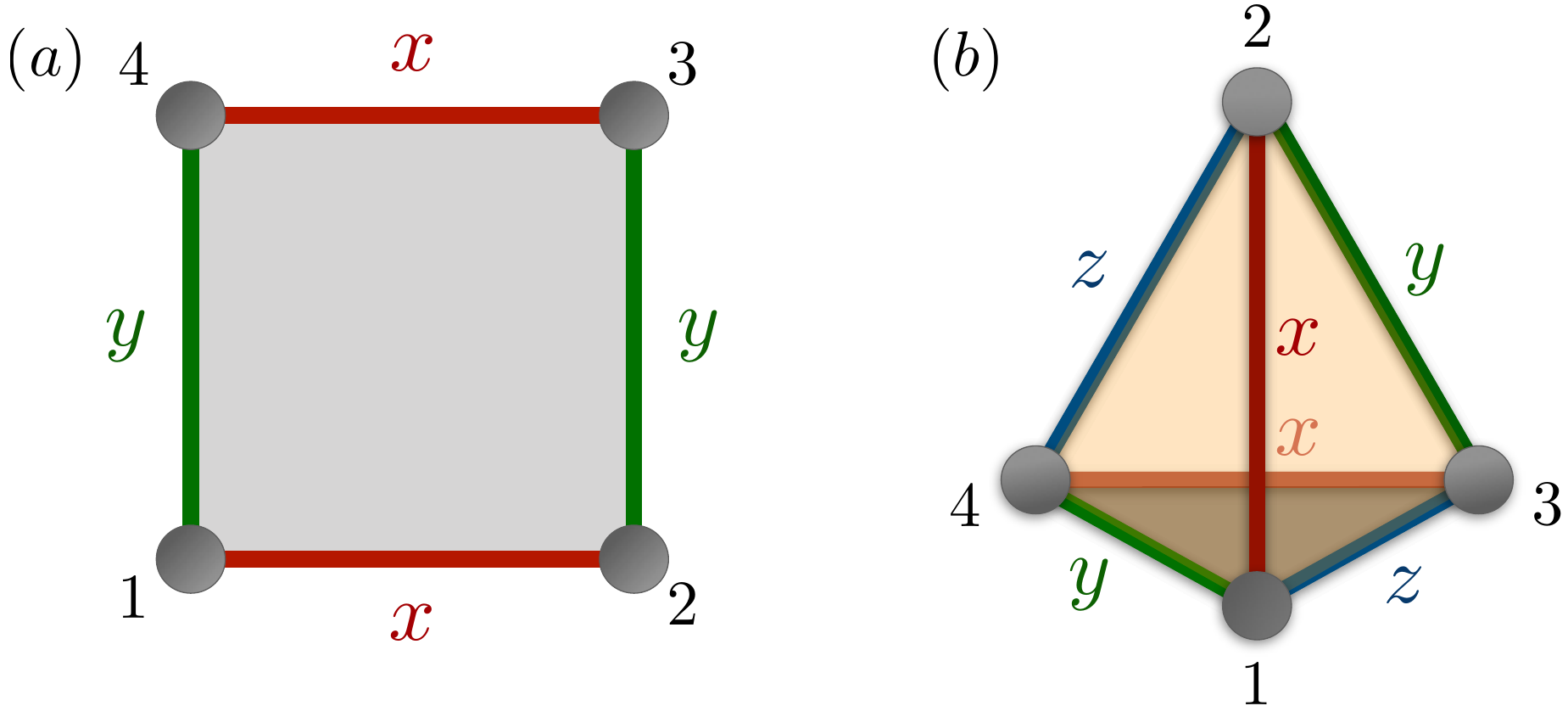}
\caption{The Kitaev square (a) and tetrahedron (b) clusters.}
\label{fig.clusters}
\end{figure}

We consider the Hamiltonian of Eq.~\ref{eq.Ham} on the square and tetrahedral clusters shown in Fig.~\ref{fig.clusters}. The tetrahedron can be obtained from the square by introducing diagonal bonds with $z-z$ couplings. We take the Kitaev coupling to be antiferromagnetic with $K>0$, without loss of generality. This can be seen by a combination of spin rotations: Keeping the spin at site 1 fixed, we rotate (i) the spin at site 2 by $\pi$ about the Z axis (i.e., $\{\hat{S}_2^x,~\hat{S}_2^y,~\hat{S}_2^z\} \rightarrow \{-\hat{S}_2^x,~-\hat{S}_2^y,~\hat{S}_2^z\} $), (ii) the spin at site 3 by $\pi$ about the Y axis (i.e., $\{\hat{S}_3^x,~\hat{S}_3^y,~\hat{S}_3^z\} \rightarrow \{-\hat{S}_3^x,~\hat{S}_3^y,~-\hat{S}_3^z\} $), and (iii) the spin at site 4 by $\pi$ about the X axis (i.e., $\{\hat{S}_4^x,~\hat{S}_4^y,~\hat{S}_4^z\} \rightarrow \{\hat{S}_4^x,~-\hat{S}_4^y,~-\hat{S}_4^z\} $). 
Rewriting the spin operators in the new rotated bases, we obtain the same Hamiltonian but with $K \rightarrow -K$. A similar transformation applies in the Kitaev model on the honeycomb lattice\cite{Rousochatzakis2015}.
The cartesian states shown in Fig.~\ref{fig.cartesian} are for the honeycomb lattice with $K>0$. They take a modified form for the $K<0$ case : starting from a dimer cover, the two spins in each dimer are aligned in parallel fashion along or opposite to the bond direction. There are two possible spin orientations for a given dimer.  
 
\begin{figure}
\includegraphics[width=\columnwidth]{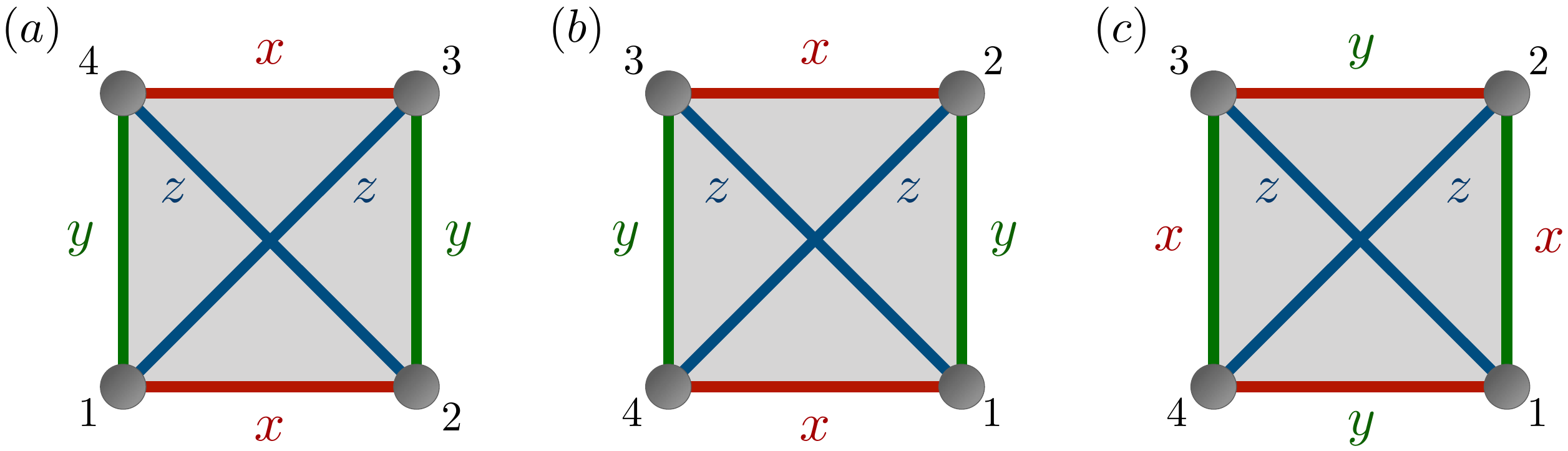}
\caption{The figure (a) shows the Kitaev tetrahedron. The Kitaev square can be obtained by simply removing the diagonal (z) bonds. The figure (b) shows the cluster after a circular permutation of the four sites. The figure (c) is obtained after a subsequent global $\pi/2$-rotation in spin space about the spin-$z$ axis.  }
\label{fig.permutation_rotation}
\end{figure}

The clusters shown in Fig.~\ref{fig.clusters} have a Hilbert space of dimension $(2S+1)^4$ with states labelled as $\vert m_1,m_2,m_3,m_4\rangle$, where $m_i$'s represent $S_z$ quantum numbers. This Hilbert space grows rapidly with $S$, placing constraints on numerical exact diagonalization. We use the following two symmetries to find the spectra: (i) The square and tetrahedron Hamiltonians are symmetric under $\pi$-rotation about the Z axis. This allows us to identify even and odd sectors, characterized by even/odd values of $m_{tot}=\sum_i m_i$. (ii) The Hamiltonians are invariant under a combination of operations: a cyclic permutation of sites followed by a global spin rotation about $\hat{z}$ by $\pi/2$. This is depicted in Fig.~\ref{fig.permutation_rotation}. Applying this symmetry four times is equivalent to an identity operation. This allows us to identify a pseudomomentum quantum number, $q=2\pi j/4$, with $j=0,1,2,3$. These two symmetries can be applied independently. We construct reduced Hamiltonian blocks by grouping together states for each $q$ value, with $m_{tot}$ restricted to either even or odd values. 

\section{Kitaev square}
\label{sec.square}
\begin{figure}
\includegraphics[width=\columnwidth]{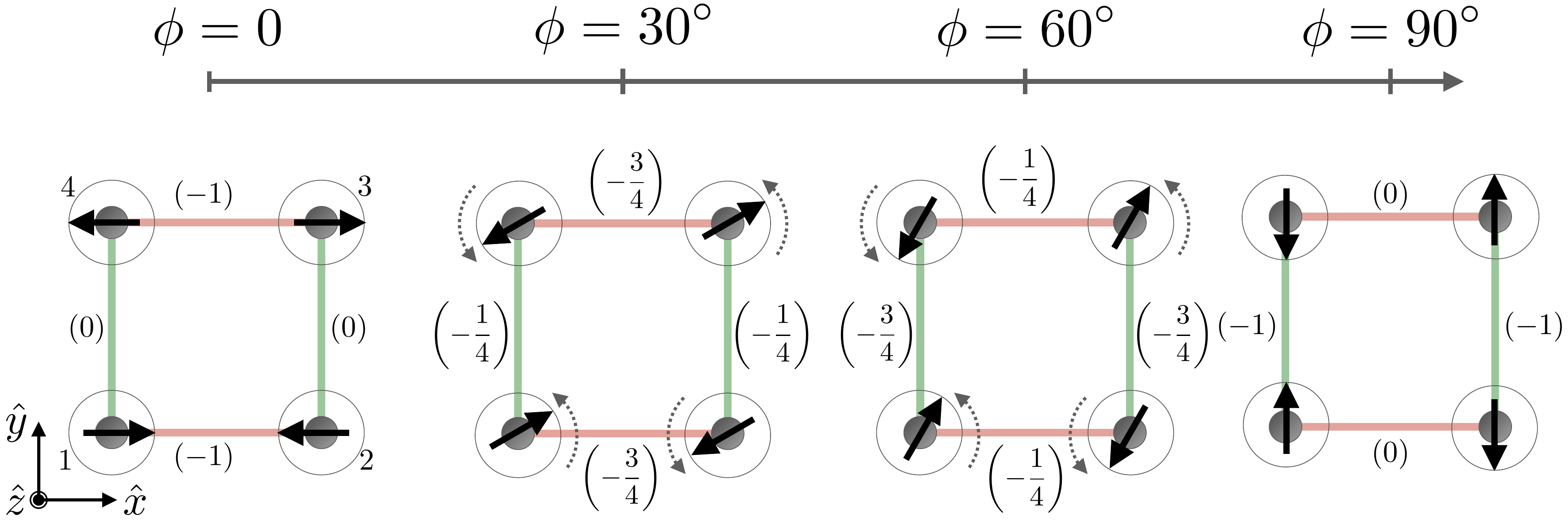}
\caption{A smooth one-parameter transformation that connects two Cartesian states. At $\phi=0$, we have a Cartesian state, corresponding to dimers on horizontal ($x$) bonds. At $\phi=90^\circ$, we have a Cartesian state corresponding to dimers on vertical ($y$) bonds. Intermediate values of $\phi$ interpolate between these states, with each spin rotating as indicated. In each state, we show the bond energies (in units of $K/S^2$) in parentheses.  }
\label{fig.deform}
\end{figure}

The Kitaev square is shown in Fig.~\ref{fig.clusters}(left). It can be viewed as a simple realization of a compass model on a four-site chain\cite{Nussinov2013}.
It can also be thought of as a one-dimensional Kitaev chain as studied by BSS in Ref.~\onlinecite{Baskaran2008}, with four sites and periodic conditions. This provides a simple starting point to understand the connectivity of classical ground states. 
BSS introduced the notion of cartesian states and showed that they are connected by smooth, energy-preserving transformations. This is illustrated in Fig.~\ref{fig.deform} on the Kitaev square, depicting a one-parameter transformation  %tuned by an angle $\phi$, 
that interpolates between two cartesian states. At the cartesian end points, the ground state energy receives contributions solely from bonds that hold dimers (in the parent dimer cover). At intermediate states, this energy is distributed among intervening bonds as well. Similar connecting pathways can exist between other pairs of Cartesian states.

The complete ground state space can be derived by applying the method of Lagrange multipliers, as shown by BSS in Ref.~\onlinecite{Baskaran2008}. Below, we describe the geometry and connectivity of the ground state space, with the explicit derivation presented in Appendix.~\ref{App.squareCGSS}.

\subsection{Space of classical ground states} 
The CGSS consists of four sectors, $C_i$ with $i=1,2,3,4$ as shown in Tab.~\ref{tab.squareCGSS}. The states in each sector are described by a free angle variable, $\phi$.  In geometric terms, each sector can be viewed as a circle. We will see below that the circles intersect at points, as in the two-circle problem discussed in Sec.~\ref{sec.twocircles} above.  
To better understand the connectivity of this space, we take these circles to be embedded in an abstract four-dimensional space with coordinates $(xyuv)$. Note that $x$, $y$, $u$ and $v$ are directions in the embedding space, distinct from directions in spin space. The distinction will be clear from context in the arguments below.

\begin{figure}
\includegraphics[width=0.75\columnwidth]{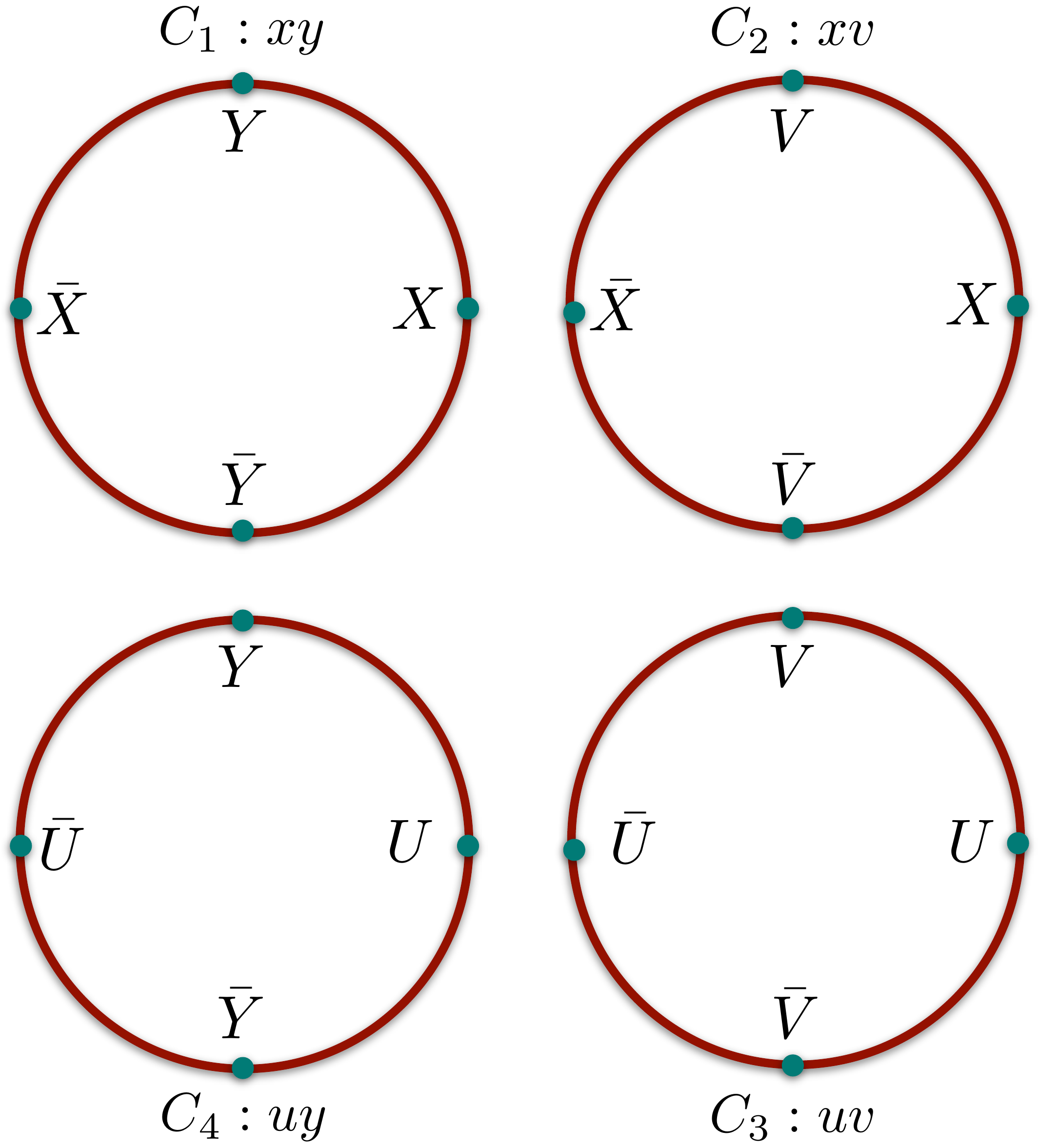}
\caption{CGSS of the Kitaev square with four circles embedded in four dimensions. Each circle lies in the plane indicated, e.g., the $C_1$ circle lies in the $(xy)$ plane.   }
\label{fig.fourcircles}
\end{figure}

To examine if the circles $C_i$ intersect, we first consider $C_1$ and $C_2$.
As can be seen from the spin configurations in Tab.~\ref{tab.squareCGSS}, a generic point in $C_1$ does not appear in $C_2$ as all the four spins $\mathbf{S}_i$ cannot be the same. However, there are two points in common, corresponding to $\phi=0$ and $\pi$ in both. We visualize the connectivity of $C_1$ and $C_2$ as shown in Fig.~\ref{fig.fourcircles}. We take $C_1$ to be a circle in the $(xy)$ plane, centred at the origin with unit radius. Each point on this circle corresponds to a certain value for the angle $\phi$, with $\phi=0$ and $\phi=\pi$ representing points that lie on the positive-$x$ and negative-$x$ axes. We represent these points as $X$ and $\bar{X}$ respectively. Similarly, we take $C_2$ to be a circle in the $(xv)$ plane with unit radius and centre at the origin. We once again take $\phi=0,\pi$ to represent points where the circle intersects the positive- and negative-$x$ axis, coinciding with $X$ and $\bar{X}$. As the two circles lie in orthogonal planes, they share a common diameter on the $x$ axis whose ends at $X$ and $\bar{X}$ constitute two points of intersection. We see that $C_1$ and $C_2$ resemble the space of two intersecting circles described in Sec.~\ref{sec.twocircles}.

Proceeding in the same manner, we take $C_3$ and $C_4$ to lie in the $(uv)$ and $(uy)$ planes respectively. The connectivity of pairs of circles can be seen in Fig.~\ref{fig.fourcircles}. For example, $C_1$ and $C_4$ intersect at two points ($Y$ and $\bar{Y}$), while $C_1$ and $C_3$ do not intersect.
An interesting geometry emerges with four circles embedded in four dimensions. Each circle intersects two other circles, sharing one common diameter with each of the them. At the same time, it remains completely distinct from the fourth circle. 

\begin{table}
\begin{tabular}{|c|c|c|c|c|c|}
\hline
$~$ & $~$ & $\mathbf{S}_1$ & $\mathbf{S}_2$ & $\mathbf{S}_3$ & $\mathbf{S}_4$ \\
\hline
$C_1$ & $xy$& $( C_\phi, S_\phi) $ & $(-C_\phi,-S_\phi)$ & $(C_\phi,S_\phi)$ & $(-C_\phi,-S_\phi)$ \\ \hline
$C_2$ & $xv$ & $( C_\phi, S_\phi) $ & $(-C_\phi,S_\phi)$ & $(C_\phi,-S_\phi)$ & $(-C_\phi,-S_\phi)$ \\ \hline
$C_3$ & $uv$ & $( C_\phi, S_\phi) $ & $(-C_\phi, S_\phi )$ & $(-C_\phi,-S_\phi)$ & $(C_\phi,-S_\phi)$  \\ \hline
$C_4$ & $uy$ & $( C_\phi, S_\phi) $ & $(-C_\phi, -S_\phi )$ & $(-C_\phi,S_\phi)$ & $(C_\phi,-S_\phi)$\\
\hline
\end{tabular}
\caption{Classical ground states of the Kitaev square. We have four families denoted by $C_i$, with $i=1,2,3,4$. States in each family are parametrized by an angle $\phi$, with $C_\phi = \cos\phi$ and $S_\phi = \sin \phi$. As all classical ground states lie in the XY plane, we only show the $(x,y)$ components for each spin.  }
\label{tab.squareCGSS}
\end{table}

\subsection{Physical interpretations of CGSS features}
\begin{figure}
\includegraphics[width=\columnwidth]{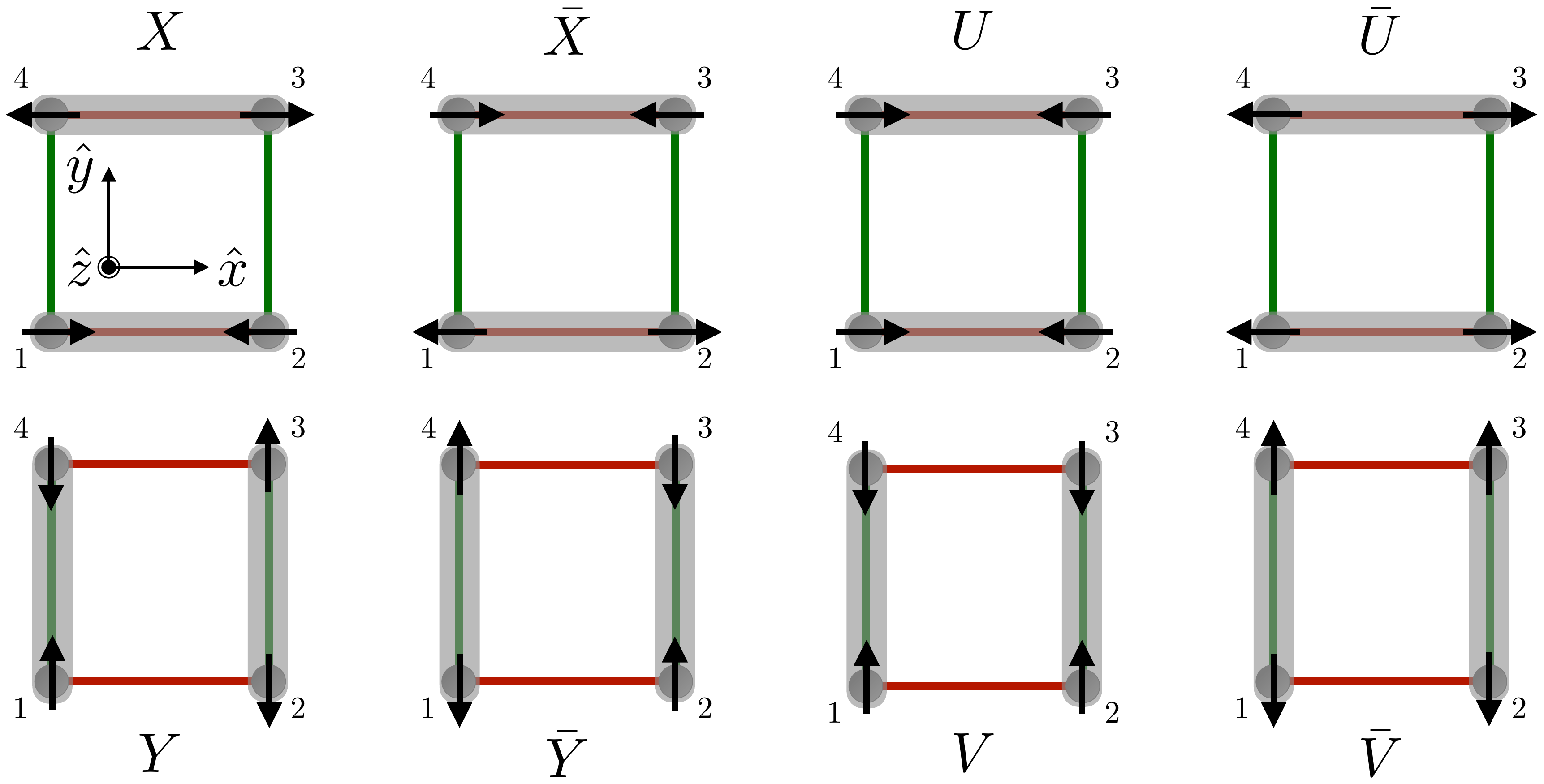}
\caption{Cartesian states on the Kitaev square.}
\label{fig.sq_cartesian}
\end{figure}

Remarkably, the points of intersection between circles are all Cartesian states. For example, $C_1$ and $C_2$ intersect when $\phi=0,\pi$ ($X$ and $\bar{X}$ respectively), corresponding to states $\{\mathbf{S}_1, \mathbf{S}_2, \mathbf{S}_3, \mathbf{S}_4\} = \pm S \{\hat{x},-\hat{x},\hat{x},-\hat{x}\}$. Here, $\phi=0$ and $\pi$ correspond to the $+$ and $-$ signs respectively. To recapitulate the definition of cartesian states, they are obtained from a dimer cover by orienting spins to maximally satisfy the bonds on each dimer. Here, these two states can be understood to emerge from a dimer cover with dimers on bonds $(1,2)$ and $(3,4)$ (see Fig.~\ref{fig.clusters}). As these bonds have $x-x$ couplings, the spins are oriented along $\pm \hat{x}$ to maximally satisfy these bonds. The intersection points $U$ and $\bar{U}$ also maximally satisfy the $x-x$ bonds, i.e., they correspond to cartesian states constructed from the same dimer cover. In contrast, the points $Y$, $\bar{Y}$, $V$ and $\bar{V}$ maximally satisfy the $y-y$ bonds. These states are shown in Fig.~\ref{fig.sq_cartesian}.

As seen in Fig.~\ref{fig.fourcircles}, pairs of cartesian states are connected by quarter arcs, e.g., $X$ and $Y$ are connected by a quarter arc in $C_1$. Such an arc represents a smooth transformation that takes us from one cartesian to another, while preserving the classical energy. This is precisely the transformation depicted in Fig.~\ref{fig.deform} above.

\subsection{Spectrum in the quantum spin-$S$ problem}
\begin{figure*}
\includegraphics[width=2\columnwidth]{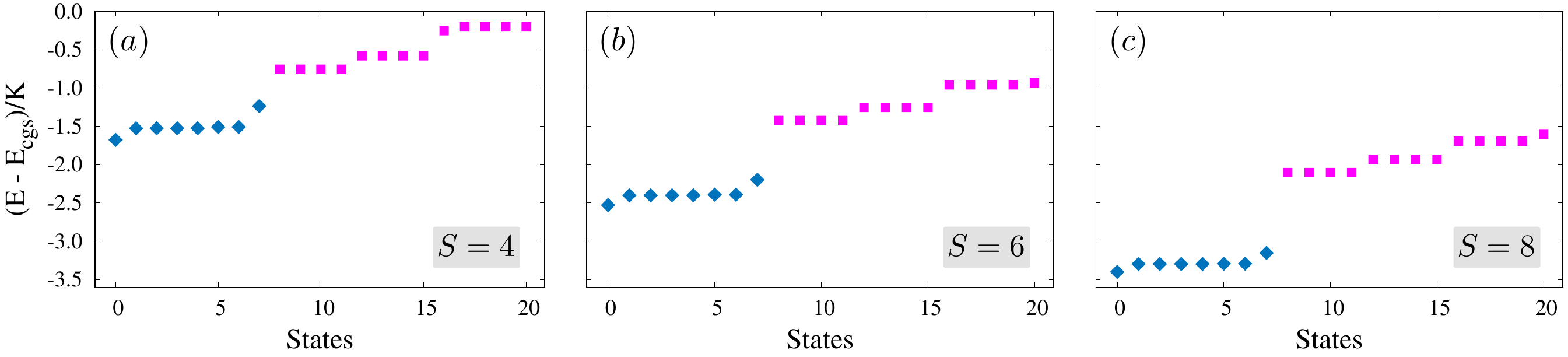}
\caption{Low energy spectra in the Kitaev square for three different spin values, $S = 4,6,8$ (a, b, c respectively). The spectra have been shifted by the classical ground state energy ($E_{cgs} = -2KS^2$) and scaled by the coupling strength, $K$. The lowest eight states are shown with blue diamonds in order to distinguish them from higher energy states (magenta squares).
}
\label{fig.sq_spectra}
\end{figure*}

We have demonstrated that the CGSS for the Kitaev square consists of four circles, with the circles intersecting at points. This can be seen as a higher dimensional generalization of the two-circle space discussed in Sec.~\ref{sec.twocircles} above. We assert that the low energy spectrum of the Kitaev square cluster maps to a particle moving on this space. We present results from numerical exact diagonalization of the spin problem for various $S$ values below. We find striking features in the low energy spectrum that can be understood in analogy with the two-circle problem. In particular, we find the aspects outlined at the end of Sec.~\ref{sec.twocircles} to hold true here.
In Fig.~\ref{fig.sq_spectra}, we show the low energy spectrum for three different values of $S$. We find a set of eight states at the bottom of the spectrum. As $S$ increases, these eight states progressively separate from the other, higher energy, states. We note that eight is precisely the number of self-intersection points in the CGSS of this problem. Equivalently, it is the number of cartesian states in the Kitaev square, as shown in Fig.~\ref{fig.sq_cartesian}. This is consistent with the insight gained in Sec.~\ref{sec.twocircles} in the two-circle problem. 

\begin{figure}
\includegraphics[width=2.5in]{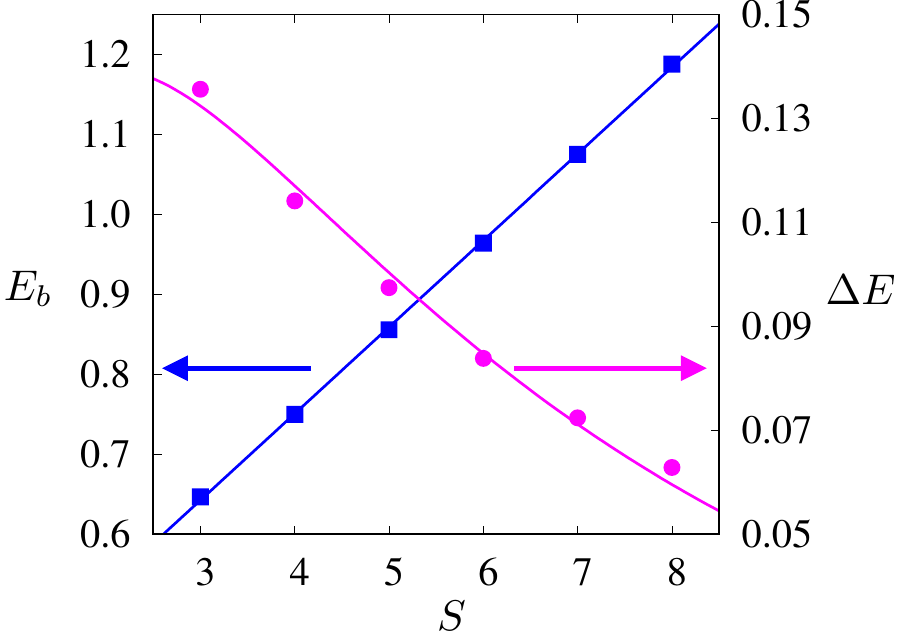}
\caption{Binding energy and spread of the low-lying set of eight eigenvalues. The binding energy is shown using blue squares. The plotted curve is the fitting function, given by $E_b (S) = 0.317375 + 0.108314 S$. The spread is shown using magenta circles. The corresponding fitting curve is given by $\Delta E(S) =0.92016\, S^{3/2}\,\exp(-2.0708\sqrt{S})$.
}
\label{fig.binding_spread}
\end{figure}

In Fig.~\ref{fig.binding_spread}, we plot two quantities that characterize the low energy spectrum. We have eight low-lying `bound' states that are separated from higher `unbound' states. To quantify the separation, we define the binding energy as $E_b = E_9 - \bar{E}$. Here, $E_9$ is the energy of the ninth state, i.e., the energy of the lowest unbound state. The average of the eight lowest states is denoted as $\bar{E}$. In Fig.~\ref{fig.binding_spread}, we see that the binding energy increases linearly with $S$. This shows that state selection becomes stronger with increasing $S$. In the $S\rightarrow\infty$ limit, all classical ground states have the same energy to $\mathcal{O}(S^2)$. However, the bound states are selected due to an $\mathcal{O}(S)$ binding energy. We define a second quantity, $\Delta E$, as the standard deviation of the lowest eight energy eigenvalues. This represents the spread in the energies of the bound states, serving as a measure of hybridization. This is comparable to the energy difference between symmetric and antisymmetric combinations of bound states in Sec.~\ref{sec.twocircles}. We find that $\Delta E$ decreases with increasing $S$, in analogy with making the discretization finer in the two-circle problem. It is well described by a fit function of the form $\Delta E(S) =0.92016\,S^{3/2}\,\exp(-2.0708\sqrt{S}) $.
 $\Delta E$ vanishes exponentially in the $S \rightarrow \infty$ limit. In this limit, we expect to have eight degenerate ground states, each corresponding to an independent bound state at a cartesian intersection point.

\subsection{Character of low lying states}
\label{ssec.weight}
\begin{figure}
\includegraphics[width=2.5in]{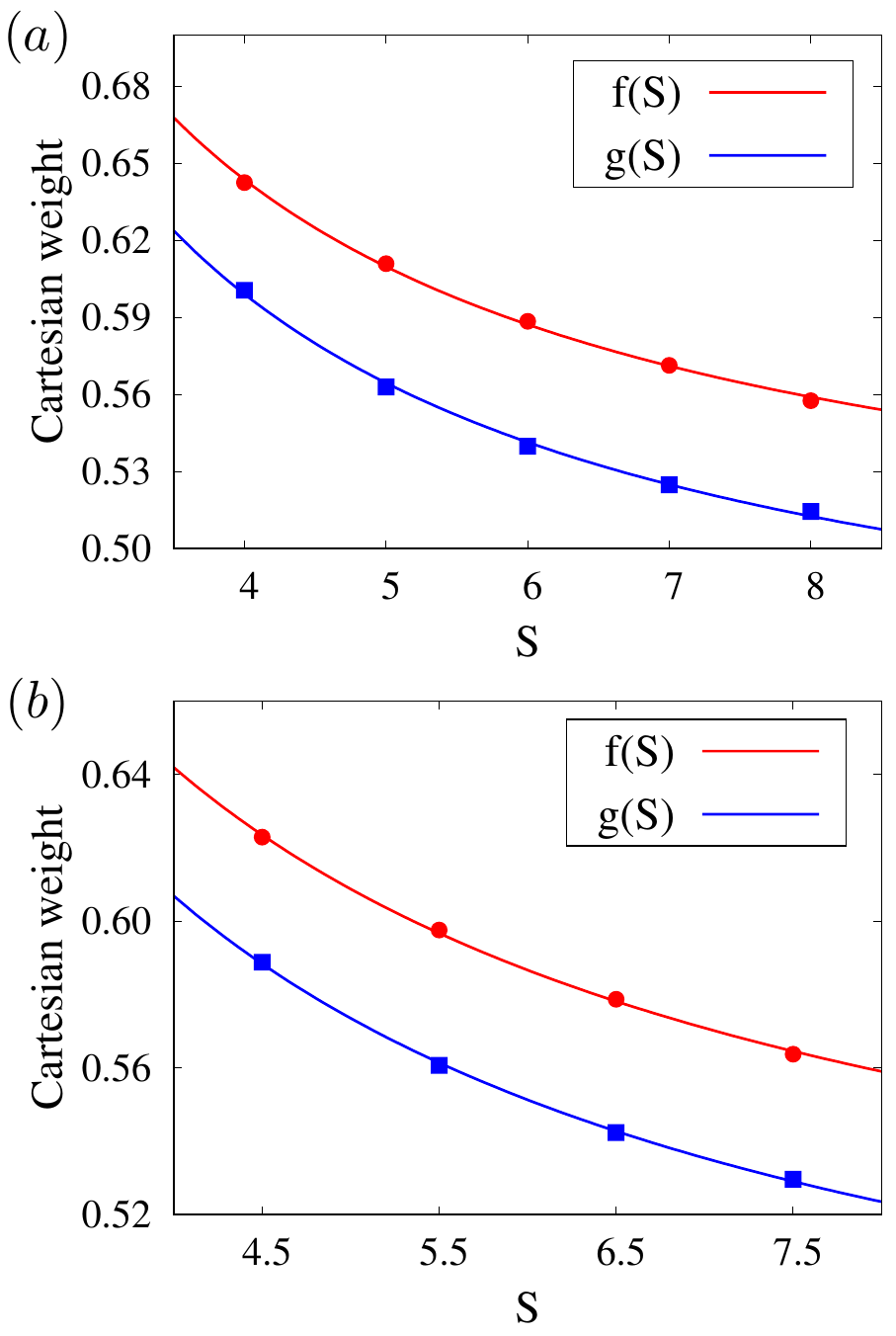}
\caption{Cartesian weight in the low-lying states of the Kitaev square for (a) integer and (b) half-integer values of $S$. The blue squares show the cartesian weight in the ground state vs. $S$. The red circles show the average cartesian weight of the eight low-lying states. For integer $S$ (top), the fitting curves are given by $f(S) = 0.474367 + 0.677225 /S$ and $g(S) = 0.426014 + 0.692372/S$. For half-integer $S$ (bottom), the fitting curves are $f(S) = 0.476134 + 0.663209/S $ and $g(S) = 0.440098 + 0.667208/S $.}
\label{fig.sq_proj}
\end{figure}

We next examine the character of the eight low-lying states, labelled as $\vert \phi_{low.}^i \rangle$, with $i=1,\ldots,8$. We surmise that these arise from bound states that are localized at self-intersection points in the CGSS. In turn, the self-intersection points correspond to cartesian states. To test this notion, we plot the `cartesian weight' in the low-lying states in Fig.~\ref{fig.sq_proj}. We calculate this as follows. 

We first adapt the classical notion of the cartesian state to the quantum spin-$S$ context. This is achieved using coherent states\cite{Auerbach_book}. For example, the cartesian state $X$ in Fig.~\ref{fig.fourcircles}, corresponding to $\{\mathbf{S}_1, \mathbf{S}_2, \mathbf{S}_3, \mathbf{S}_4\} =S \{\hat{x},-\hat{x},\hat{x},-\hat{x}\}$, is written as
\begin{eqnarray}
\vert C_X \rangle = 
\exp \big[ {-i\frac{\pi}{2} \{ \hat{S}_1^y - \hat{S}_2^y +\hat{S}_3^y -\hat{S}_4^y \}} \big] \vert S,S,S,S\rangle. 
\label{eq.cartX}
\end{eqnarray}
Here, $\vert S,S,S,S \rangle$ is the state with all spins polarized along $\hat{z}$. We write the seven other cartesian states in the same fashion. We seek to quantify the contribution of these eight cartesian states to the eight low-lying states in the spectrum.

However, a subtlety arises here as the cartesian states are not mutually orthogonal. For example, 
$\langle C_X \vert C_Y \rangle \neq0$, where $\vert C_Y\rangle$ corresponds to $\{\mathbf{S}_1, \mathbf{S}_2, \mathbf{S}_3, \mathbf{S}_4\}  =S\{\hat{y},-\hat{y},\hat{y},-\hat{y}\}$. 
In order to disentangle these states, we use a Gram-Schmidt procedure to find $\vert \Phi_j \rangle$, $j=1,\ldots,8$, a set of eight mutually orthogonal states that are linear combinations of cartesian states. These span an eight-dimensional subspace of the full Hilbert space. We define the `cartesian weight' of a low-lying state as its weight in this subspace, given by $P_i = \sum_{j=1}^8 \vert\langle \Phi_j \vert \phi_{low.}^i \rangle \vert^2$. If $\vert \phi_{low.}^i \rangle$ has no contributions from cartesian states, $P_i$ would be zero. In contrast, if it is composed entirely of cartesian states, $P_i$ would be unity. Based on the results of the two-circle problem in Sec.~\ref{sec.twocircles}, we expect the cartesian weight of the eight low energy states to be finite and less than unity. 
As seen in Fig.~\ref{fig.twocircles}, the two lowest states in the two-circle problem are indeed localized at the intersection points. However, they are not singularly localized with delta-function-like nature. Rather, they decay exponentially with the strongest amplitude at the cartesian points. We expect the eight lowest states in the Kitaev square to be of the same nature. We expect them to have a significant fraction of their weight contributed from cartesian states, but not their entire weight. 

Our results for the cartesian weight are shown in Fig.~\ref{fig.sq_proj} for various $S$ values. The figure plots two quantities. The first is the cartesian weight of the ground state, i.e., the lowest of the eight low-lying states. The second is the average cartesian weight over all eight low-lying states. In both cases, the cartesian weight is a significant fraction, e.g., the $S=5$ ground state has a $\sim$56$\%$ contribution from the cartesian states.  We emphasize that this represents a very large contribution. The cartesian states are only eight elements in the Hilbert space of size $11^4 = 14641$. Yet, these eight states carry more than half the weight of the ground state. From Fig.~\ref{fig.sq_proj}, we note that the cartesian contribution in the ground state is always less than the average cartesian weight over all eight states. This can be understood by analogy with the two-circle problem. The ground state there is a symmetric combination of bound states at the two intersection points. It has significant weight in the  intermediate regions due to constructive interference. As a consequence, the weight at the intersection points is somewhat diminished. In contrast, the first excited state, being an anti-symmetric combination, has a larger weight at the intersection points. In the same manner, we believe that the ground state of the Kitaev square is a symmetric combination of bound states. As a result, it has a smaller cartesian weight than the other seven states.  

The $S$-dependence of the cartesian weight can be seen in Fig.~\ref{fig.sq_proj}. We find a smooth evolution with $S$ if we separate integer and half-integer values of $S$ as shown. In Appendix.~\ref{App.Berry}, we demonstrate that a non-trivial Berry phase emerges that distinguishes these two cases. This is in line with arguments presented in Ref.~\onlinecite{Khatua2019}. The spin problem maps to that of a single particle moving on the CGSS. When the spin system evolves along a closed path in the CGSS space, it can accrue a Berry phase. This is a well known ingredient in the spin path integral formulation. In the mapping to the single particle picture, this translates to an Aharonov-Bohm phase that can alter the spectrum. Here, there is a path within the CGSS which accrues a Berry phase when $S$ is a half-integer, but not when $S$ is an integer.
The $S$-dependence is captured by polynomial fits to the data as shown in  Fig.~\ref{fig.sq_proj}. From the fit functions, we surmise that the cartesian weight extrapolates to a non-zero value as $S\rightarrow\infty$. As we approach this limit, the number of cartesian states remains fixed at eight while the Hilbert space size grows exponentially. Despite this, the cartesian states retain a finite weight at $S\rightarrow \infty$. 

In summary, the low energy physics of the Kitaev square is dominated by cartesian states. We see this in the spectrum as a set of eight low-lying states, energetically separated from all other states. These states are, in fact, quantum analogues of the classical cartesian states. The energy gap to other states increases with increasing $S$, indicating that cartesian states determine the low energy behaviour in the classical $S\rightarrow\infty$ limit. We provide an independent test of these results in the following subsection. 

\subsection{Cartesian fidelity of the low-lying states}
\label{ssec.fidelity}
We have argued that the lowest energy states of the Kitaev square are essentially admixtures of the eight cartesian states. It follows that we can recover the cartesian states by suitably mixing the low energy states. In order to achieve this, we define a resolving operator,
\begin{eqnarray}
\nonumber \hat{O}_{sq,res.} &=& \lambda_{12} (\hat{S}_1^x - \hat{S}_2^x  ) + \lambda_{34}  (\hat{S}_3^x - \hat{S}_4^x  ) \\
&+& \lambda_{14}  (\hat{S}_1^y - \hat{S}_4^y  ) + \lambda_{23}  (\hat{S}_2^y - \hat{S}_3^y  ), 
\label{eq.Osq}
\end{eqnarray}
where the $\lambda$ coefficients are chosen to be substantially different from one another. We have one coefficient for each bond, linearly coupling to the Ising antiferromagnetic moment along the bond direction. For example, the bond $(1,2)$ has an $x-x$ coupling in the Kitaev square Hamiltonian. We have one term associated with this bond in $\hat{O}_{sq,res.}$, given by $ \lambda_{12} (\hat{S}_1^x - \hat{S}_2^x  )$. This term serves as a diagnostic for cartesian states in the following manner. 
We evaluate its expectation value in a cartesian state, i.e., in the quantum spin-$S$ version of a cartesian state.
If the cartesian state has a dimer on this bond, this term contributes $ \pm 2\lambda_{12} S$, with the $+$ or $-$ sign depending on the orientations of spins on this dimer. In a cartesian state which does not have a dimer on this bond, this term has expectation value zero. Thus, this term resolves two specific cartesian states. In the same way, each term in $\hat{O}_{sq,res.}$ serves as an indicator for two cartesian states.

Our premise is that the eight low-lying states in the spin-$S$ Kitaev square problem are essentially composed of cartesian states. We test this notion by examining the expectation values of the resolving operator in the low-lying states. We find its matrix elements, $O_{mn} = \langle \phi_{low.}^m \vert \hat{O}_{sq,res.} \vert \phi_{low.}^n \rangle$. We now diagonalize the $8 \times 8$ matrix that has $O_{mn}$ as its entries. We find that its eight eigenvalues are approximately given by $(\pm \lambda_{12} \pm \lambda_{34}),~ (\pm \lambda_{14} \pm \lambda_{23})$. These expressions correspond precisely to the expectation values of $\hat{O}_{sq,res.}$ in the eight cartesian states. This shows that the eight low-lying states can be mixed with one another to realize the cartesian states. Note that the low-lying states span an eight-dimensional subspace, as do the cartesian states. We proceed to define a single parameter that quantifies the equivalence between them.

\begin{figure}
\includegraphics[width=2.5in]{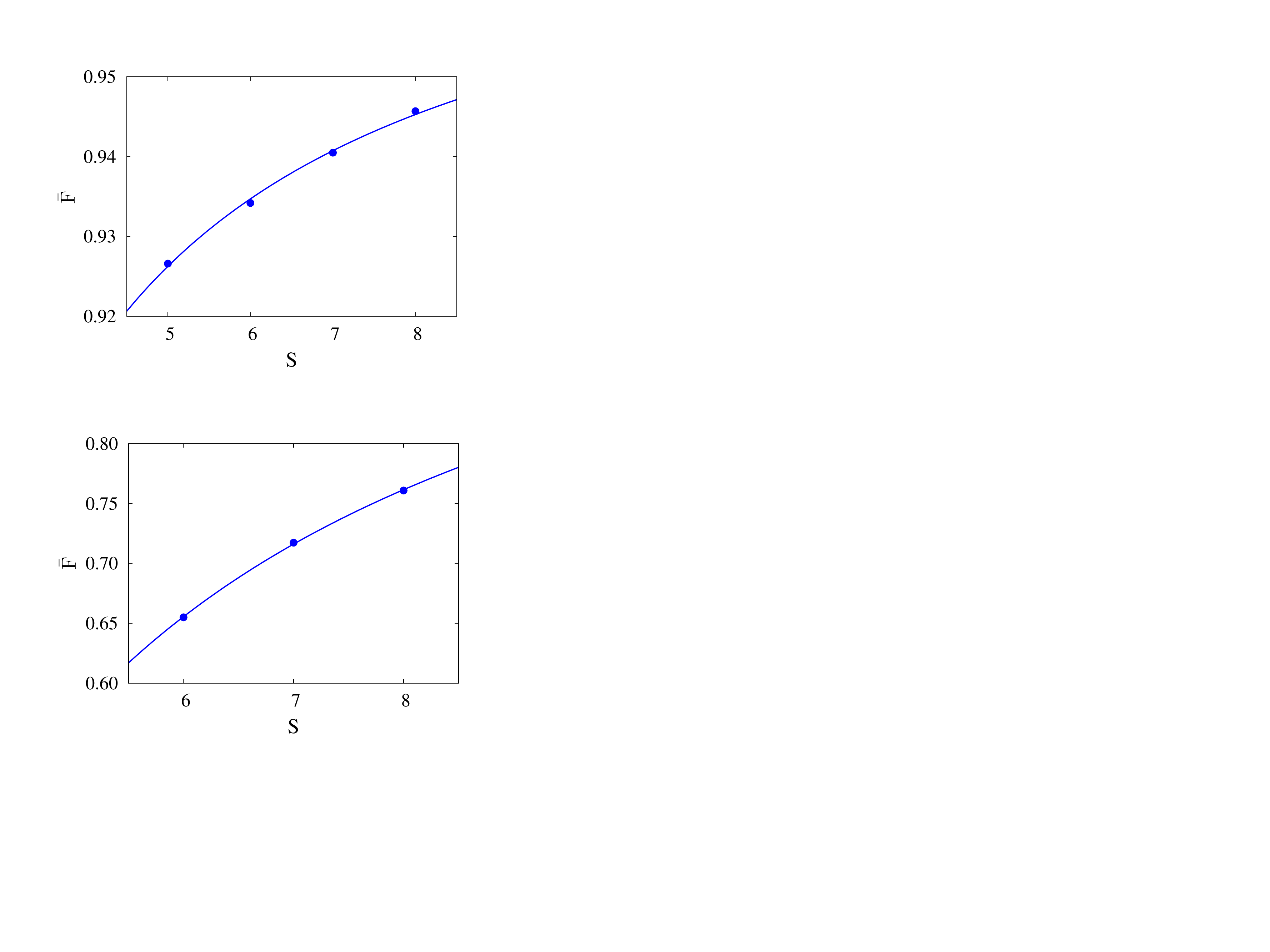}
\caption{Cartesian fidelity in the Kitaev square as a function of $S$. The data is fit using $\bar{F}(S) = 0.97699 - 0.25367/S $. }
\label{fig.sq_fidelity}
\end{figure}

As noted above, the eigenvalues of $O_{mn}$ are close to expectation values of $\hat{O}_{sq,res.}$ in the cartesian states. As the $\lambda$'s in Eq.~\ref{eq.Osq} are chosen to significantly differ from one another, we can clearly distinguish the eigenvalues and identify them with corresponding cartesian states. 
This establishes a one-to-one relationship between the \textit{eigenstates} of $\hat{O}_{sq,res.}$ (mixtures of the eight low-lying states) and cartesian states. 
We label the eigenstates as $\vert \ell_{\alpha} \rangle$, with $\alpha=1,\ldots,8$. We express the associated cartesian states as $\{ \mathbf{S}_{\alpha,j} , j=1,\ldots,4\}$, denoting the (classical) orientation of the $j^{th}$ spin in the $\alpha^{th}$ cartesian state.
To quantify the fidelity of this relationship, we define vectors $\mathbf{v}_{\alpha,j}$ as follows. 
Here, $\alpha=1,\ldots,8$ identifies one of the eigenvectors of $O_{mn}$ while $j=1,\ldots,4$ represents one of the four sites in the Kitaev square. We define $ \mathbf{v}_{\alpha,j} \equiv \langle \ell_\alpha \vert \hat{\mathbf{S}}_j \vert \ell_\alpha \rangle$. As each $\vert \ell_\alpha \rangle$ maps to one particular cartesian state, we find that each vector $\mathbf{v}_{\alpha,j}$ closely resembles the spin configuration of a cartesian state. The fidelity of this mapping is seen by defining a quantity,
\begin{eqnarray}
F_\alpha =\frac{1}{4 S^2} \sum_{j=1}^4 \mathbf{v}_{\alpha,j} \cdot \mathbf{S}_{\alpha,j}.
\label{eq.Falpha}
\end{eqnarray}
If the eight low-lying states were composed purely of cartesian states, the states $\vert \ell_\alpha \rangle$ would be precisely the cartesian states. This would be reflected in the spin expectation values, with $\mathbf{v}_{\alpha,j}= \mathbf{S}_{\alpha,j} $. The quantity $F_\alpha$ would then take its maximum value of unity. In practice, we expect the low-lying states to not just be composed of cartesian states, but to have some additional contributions from nearby states. For example, in the particle picture, the bound state wavefunctions also have non-zero contributions from points that are close to the self-intersection points. As a consequence, we expect $F_\alpha$ to be generically less than unity. In Fig.~\ref{fig.sq_fidelity}, we plot $\bar{F}$, i.e., $F_\alpha$ averaged over all $\alpha$, as a function of $S$. We see that $\bar{F}$ increases with $S$ and, more importantly, approaches unity as $S\rightarrow \infty$. This indicates that the eight low-lying states are indeed essentially composed of cartesian states. Their cartesian character increases with increasing $S$.

\subsection{Scaling relations in the spectrum}
\label{ssec.scaling}
We have demonstrated that the eight low-lying states correspond to cartesian states. In turn, these correspond to self-intersection points in the CGSS. In Fig.~\ref{fig.binding_spread}, we have described two quantities, the binding energy and the spread. The former increases linearly with $S$, while the latter decreases with $S$. We now rationalize these empirical observations with suitable scaling arguments.

The low energy physics of the Kitaev square maps to a single particle moving on the CGSS. We model the dynamics of the particle using a tight binding description. This involves two parameters: the hopping amplitude $t$ as well as the arc length $L$. The latter represents the density of the tight binding mesh. The CGSS consists of line segments that connect cartesian points, e.g., we have an arc that connects the points $X$ and $Y$ within $C_1$, as seen in Fig.~\ref{fig.fourcircles}. In the tight binding scheme, we take this arc to consist of $L$ sites. The evolution of the spectrum with $S$ is encoded in the tight binding parameters. We now argue that these parameters scale with $S$ in a characteristic manner with $t\sim S$ and $L\sim \sqrt{S}$. 

The CGSS is a generalization of the two-circle problem presented above. While the CGSS for the Kitaev square is bigger, the nature of the self-intersections is precisely the same. Within the tight binding scheme, both cases lead to bound states with a decay constant, $\alpha$. This quantity is independent of $t$ and $L$. It depends solely on the connectivity of the CGSS at the self-intersection point. As a consequence, we expect $\alpha$ to be independent of $S$. As argued in Sec.~\ref{sec.twocircles}, when the self-intersection points are well separated, we have bound states with binding energy $(4/\sqrt{3}-2)t$. This is the difference in energy between the bound state and the lowest delocalized state. This quantity is directly proportional to $t$ and is independent of $L$. As we empirically find that the binding energy scales as $S$ (to leading order, see Fig.~\ref{fig.binding_spread}), we conclude that $t$ scales linearly with $S$.

To determine the scaling of $L$ with $S$, we appeal to the example of the XY dimer discussed in Ref.~\onlinecite{Khatua2019}. The corresponding CGSS is a circle, resembling the CGSS of the Kitaev square at generic points. The low energy physics of the dimer maps to a particle on a circle, which can be encoded as a tight binding model. The resulting parameters, $t$ and $L$, must scale in the same way with $S$ as in the case of the Kitaev square. The spectrum of the XY dimer was explicitly worked out in Ref.~\onlinecite{Khatua2019} (see Fig.~3 therein). It was shown that it maps to a particle on a ring. In particular, the low energy states have energies given by $\epsilon\sim a_0 m^2$, where $m$ is an integer. The scale factor $a_0$ represents the inverse of the moment of inertia of the particle. Crucially, we find that $a_0$ is an $\mathcal{O}(S^0)$ quantity (to leading order in $S$). In the tight binding description, $a_0$ corresponds to the ratio $t/L^2$. As we have argued that $t \sim S$ and $a_0 \sim S^0$, we arrive at $L \sim \sqrt{S}$.

In the Kitaev square, we have bound states that form at the eight self-intersection points. These states hybridize among themselves. The spread in their energies is proportional to the overlap between bound states centred at the ends of an arc, 
\begin{eqnarray}
E_{overlap}  \sim \langle   \psi_0\vert \hat{H}_t \vert \psi_{L} \rangle \sim E_{bound} \sum_{n=0}^L e^{-\alpha n} e^{-\alpha (L-n)}.  
\end{eqnarray}
Here, the `$\sim$' sign indicates proportionality upto constants that are independent of $S$. We have bound states, $ \vert \psi_0 \rangle$  and $ \vert \psi_L \rangle$, localized at the ends of the arc. The operator $\hat{H}_t$ represents the hopping Hamiltonian on the arc. To a good approximation, $\vert \psi_L \rangle$ is an eigenstate of $\hat{H}_t$ with eigenvalue $E_{bound}$. Here, $E_{bound}$ is the energy of a bound state at a well separated self-intersection point, as derived in Sec.~\ref{sec.twocircles}. It is proportional to the hopping amplitude, $t$. In evaluating the overlap, we have used the explicit form of the bound state wavefunction given in Eq.~\ref{eq.bound}. For simplicity, we have assumed that the overlap only receives contributions from sites on the intervening arc, denoted by the index $n$. The contributions from sites on other arcs will be negligible. 
We find $E_{overlap} \sim E_{bound} L e^{-L \alpha}$. Using the scaling relations for $t$ and $L$, we have $E_{overlap} \approx a S^{3/2} \exp(-b\sqrt{S})$. In Fig.~\ref{fig.binding_spread}, we have fit $\Delta E$ to this functional form, obtaining $a$ and $b$ as fitting parameters. 

We have argued that parameters in the effective tight binding model scale as $t\sim S$ and $L\sim \sqrt{S}$. These scaling relations are consistent with the numerically obtained spectrum in the Kitaev square problem. In particular, they provide a rationalization for the binding energy increasing linearly with $S$.
These scaling relations may be more general applicable. We find that they are broadly consistent with the spectrum of the XY quadrumer, where $t$ and $L$ for a suitable tight binding model were found as fitting parameters\cite{Khatua2019}.
 
\section{Kitaev tetrahedron}
\label{sec.tetrahedron}
We now move to the Kitaev tetrahedron that has $z-z$ couplings in addition to those present in the Kitaev square. We first describe the classical ground state space of this problem, before discussing its spectrum.  

\begin{figure*}
\includegraphics[width=1.5\columnwidth]{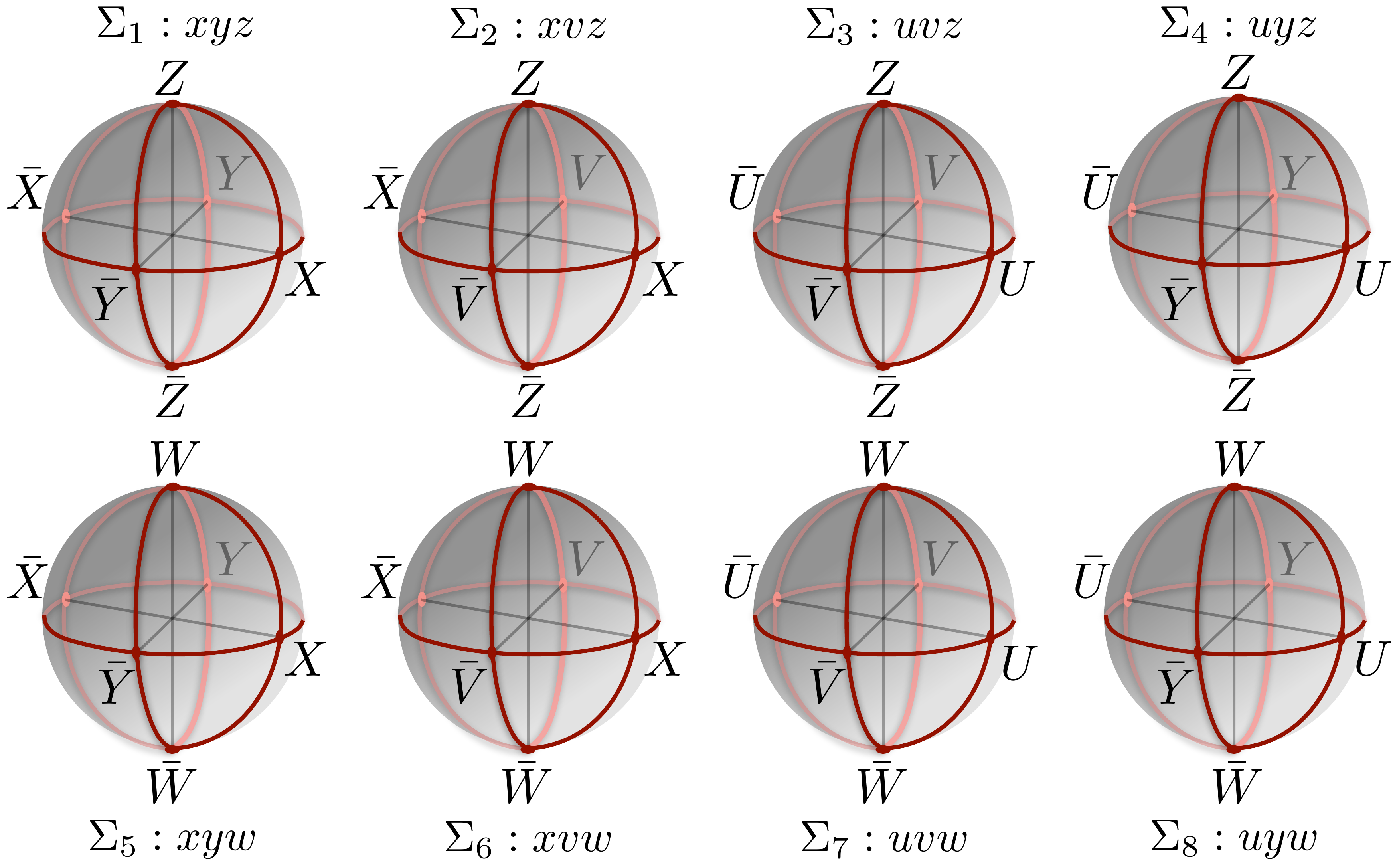}
\caption{CGSS of the Kitaev tetrahedron with eight spheres embedded in six dimensions. Each sphere lies in a three-dimensional subspace as indicated. For example, the $\Sigma_1$ sphere lies in the space spanned by $x$, $y$ and $z$ coordinates. }
\label{fig.eightspheres}
\end{figure*}

\subsection{Classical ground state space}
As with the Kitaev square, the method of Lagrange multipliers can be used to find the conditions necessary for achieving a classical ground state. We present details about energy minimization in App.~\ref{App.tetCGSS} and the resulting classical ground state framework in App.~\ref{App.tetCGSS_b}. We now proceed to describe the CGSS and its connectivity here.

\begin{table*}
    \begin{tabular}{|c|c|c|c|c|c|}
    \hline
    $~$ & $~$ & $\mathbf{S}_1$ & $\mathbf{S}_2$ & $\mathbf{S}_3$ & $\mathbf{S}_4$ \\
    \hline
    $\Sigma_1$ & ${xyz}$ & $( S_\theta C_\phi,S_\theta S_\phi,C_\theta) $ & $( -S_\theta C_\phi,-S_\theta S_\phi,C_\theta)$ & $( S_\theta C_\phi,S_\theta S_\phi,-C_\theta)$ & $( -S_\theta C_\phi,-S_\theta S_\phi,-C_\theta)$ \\ \hline
    $\Sigma_2$ & ${xvz}$ & $( S_\theta C_\phi,S_\theta S_\phi,C_\theta) $ & $( -S_\theta C_\phi,S_\theta S_\phi,C_\theta)$ & $( S_\theta C_\phi,-S_\theta S_\phi,-C_\theta)$ & $( -S_\theta C_\phi,-S_\theta S_\phi,-C_\theta)$ \\ \hline
    $\Sigma_3$ & ${uvz}$ & $( S_\theta C_\phi,S_\theta S_\phi,C_\theta) $ & $( -S_\theta C_\phi,S_\theta S_\phi,C_\theta)$ & $( -S_\theta C_\phi,-S_\theta S_\phi,-C_\theta)$ & $( S_\theta C_\phi,-S_\theta S_\phi,-C_\theta)$ \\ \hline
    $\Sigma_4$ & ${uyz}$ & $( S_\theta C_\phi,S_\theta S_\phi,C_\theta) $ & $( -S_\theta C_\phi,-S_\theta S_\phi,C_\theta)$ & $( -S_\theta C_\phi,S_\theta S_\phi,-C_\theta)$ & $( S_\theta C_\phi,-S_\theta S_\phi,-C_\theta)$ \\ \hline
    $\Sigma_5$ & ${xyw}$ & $( S_\theta C_\phi,S_\theta S_\phi,C_\theta) $ & $( -S_\theta C_\phi,-S_\theta S_\phi,-C_\theta)$ & $( S_\theta C_\phi,S_\theta S_\phi,-C_\theta)$ & $(-S_\theta C_\phi,-S_\theta S_\phi,C_\theta)$ \\ \hline
    $\Sigma_6$ & ${xvw}$ & $( S_\theta C_\phi,S_\theta S_\phi,C_\theta)$ & $( -S_\theta C_\phi,S_\theta S_\phi,-C_\theta)$ & $( S_\theta C_\phi,-S_\theta S_\phi,-C_\theta)$ & $( -S_\theta C_\phi,-S_\theta S_\phi,C_\theta)$ \\ \hline
    $\Sigma_7$ & ${uvw}$ & $( S_\theta C_\phi,S_\theta S_\phi,C_\theta)$ & $( -S_\theta C_\phi,S_\theta S_\phi,-C_\theta)$ & $( -S_\theta C_\phi,-S_\theta S_\phi,-C_\theta)$ & $( S_\theta C_\phi,-S_\theta S_\phi,C_\theta)$ \\ \hline
    $\Sigma_8$ & ${uyw}$ & $( S_\theta C_\phi,S_\theta S_\phi,C_\theta)$ & $( -S_\theta C_\phi,-S_\theta S_\phi,-C_\theta)$ & $( -S_\theta C_\phi,S_\theta S_\phi,-C_\theta)$ & $( S_\theta C_\phi,-S_\theta S_\phi,C_\theta)$ \\ \hline
    \end{tabular}
    \caption{Classical ground states of the Kitaev tetrahedron. We have eight families denoted by $\Sigma_i$, with $i=1,\ldots, 8$. States in each family are parametrized by two angles, $\theta$ and $\phi$, with $C_{\theta/\phi} = \cos(\theta/\phi)$ and $S_{\theta/\phi} = \sin (\theta/\phi)$.  }
    \label{tab.tetCGSS}
    \end{table*}
Unlike the Kitaev square, the tetrahedron also possesses non-coplanar classical ground states. By systematically analyzing the ground state conditions, we account for all ground states using two continuous variables and eight discrete choices. We thus have a CGSS composed of eight sectors, $\Sigma_i$ with $i=1,\ldots,8$, as shown in Tab.~\ref{tab.tetCGSS}. Each sector is parametrized by two angles, $\theta$ and $\phi$. These angles describe the orientation of the first spin, $\mathbf{S}_1$, in standard spherical coordinates. As this suggests, these angles satisfy the periodicity of a sphere, e.g., $\phi \equiv \phi+2\pi$. The orientations of the remaining three spins vary from sector to sector as shown in Tab.~\ref{tab.tetCGSS}. From these arguments, we deduce that each sector represents a two-sphere (${S}^2$), parametrized by the two angles $\theta$ and $\phi$. We thus have eight spheres as the CGSS. As we will see below, these spheres are not distinct as they intersect with one another. We will describe the space by suitably adapting the arguments from the Kitaev square case discussed above. 

To describe the connectivity of the space, we take the spheres to be embedded in an abstract six-dimensional space with coordinates $(xyzuvw)$. The labels $x$, $y$, etc. represent directions in the embedding space and not in spin space. We first consider the sector $\Sigma_1$ as described in Tab.~\ref{tab.tetCGSS}. Each element in this sector corresponds to a choice of $(\theta,\phi)$. We visualize this as a unit sphere in the subspace spanned by the $x$, $y$ and $z$ coordinates, i.e., as the set of points satisfying $(x^2 + y^2 + z^2 = 1;~u=v=w=0)$. The angles, $\theta$ and $\phi$, parametrize this spherical surface. We take $\theta$ to be the polar angle, measured from the $z$ axis. The azimuthal angle, $\phi$, is taken to be measured from the $x$ axis. For example, $(\theta=\pi/2,\phi=\pi/2)$ corresponds to the point $(x,y,z,u,v,w)=(0,1,0,0,0,0)$. The $\Sigma_1$ sphere is shown at top left in Fig.~\ref{fig.eightspheres}. The figure shows three great circles where the sphere intersects the $xy$, $yz$ and $zx$ planes. We will see below that these great circles have an interesting physical interpretation.

In the same manner, we represent the $\Sigma_{2,\ldots,8}$ sectors with spheres. Each sphere lies in the subspace formed by three coordinates, as indicated in Tab.~\ref{tab.tetCGSS} and shown in Fig.~\ref{fig.eightspheres}. In $\Sigma_{1,\ldots,4}$, we take the polar angle to be measured from the $z$ axis. As $\Sigma_{5,\ldots,8}$ do not extend into the $z$ direction, we measure the polar angle from the $w$ direction. In the same manner, in $\Sigma_{1}$, $\Sigma_2$, $\Sigma_5$ and $\Sigma_6$, we measure the azimuthal angle from the $x$ direction. In the remaining four, we measure it from the $u$ direction. 

Crucially, the spheres intersect one another. The geometry is much more complex than in the Kitaev square CGSS with two types of intersections: one-dimensional and zero-dimensional. To give an example of a one-dimensional intersection, we consider $\Sigma_1$, which lies in $(xyz)$ subspace, and $\Sigma_5$ which resides in $(xyw)$ subspace. These two spheres overlap along a great circle that lies in the $xy$ plane.
There are many other such one-dimensional intersections, e.g., $\Sigma_1$ (xyz) and $\Sigma_2$ (xvz) overlap along a great circle in the $zx$ plane. We call these `one-dimensional' as the locus of intersection is a circle. 
In contrast, we have a separate class of intersections that are zero-dimensional. For example, we take $\Sigma_1$ (xyz) and $\Sigma_3$ (uvz). These two spheres share a common diameter along the $z$ direction. They intersect at precisely two points given by $(x,y,z,u,v,w) = (0,0,\pm 1, 0,0,0)$. The locus of intersection here is composed of distinct `zero-dimensional' points. There are several such zero-dimensional intersections as can be seen in Fig.~\ref{fig.eightspheres}. We also have pairs of spheres that do not intersect, e.g., $\Sigma_1$ (xyz) and $\Sigma_7$ (uvw) do not have any points in common. 

\subsection{Physical interpretation of CGSS features}
\begin{figure}
\includegraphics[width=\columnwidth]{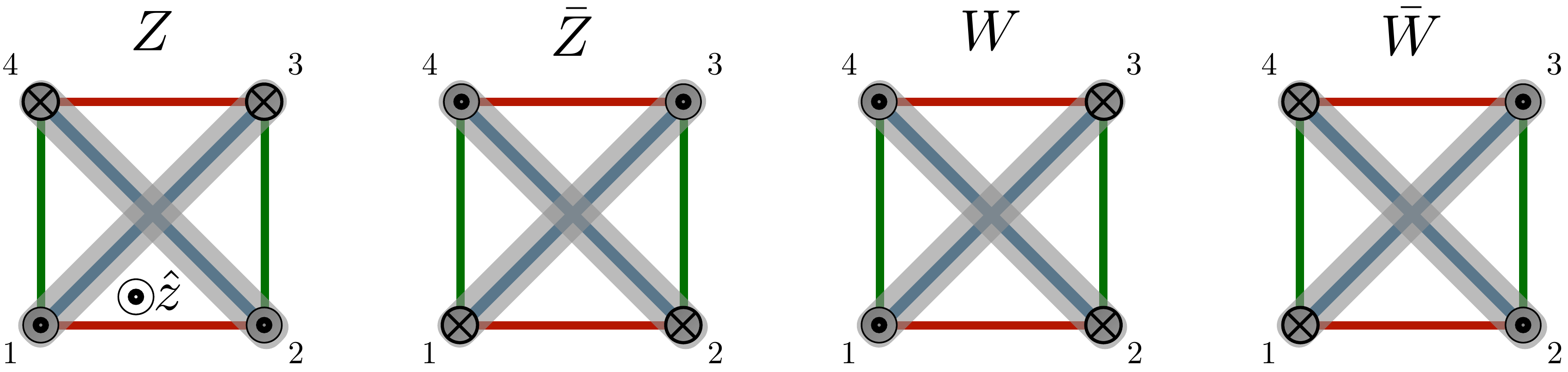}
\caption{Additional cartesian states that emerge in the Kitaev tetrahedron.}
\label{fig.ttet_cartesian}
\end{figure}

The zero-dimensional singularities in this space occur along the six cardinal directions of the embedding space. They are marked as $X$, $\bar{X}$, $Y$, $\bar{Y}$, $Z$, $\bar{Z}$, $U$, $\bar{U}$, $V$, $\bar{V}$, $W$ and $\bar{W}$ in Fig.~\ref{fig.eightspheres}. Note that there are twelve such points. These points have a remarkable interpretation in the physical spin problem: they correspond to cartesian states on the tetrahedron. For example, the point $X$ corresponds to $\{ \mathbf{S}_1, \mathbf{S}_2, \mathbf{S}_3, \mathbf{S}_4 \} = S \{ \hat{x}, -\hat{x}, \hat{x}, -\hat{x}\}$. This corresponds to a dimer cover with dimers placed on $(12)$ and $(34)$ bonds. The spins on the dimers have been aligned along $\hat{x}$ and $-\hat{x}$ so as to maximally satisfy these bonds. A simple analysis shows that there are only three possible dimer covers on the tetrahedron. With each dimer cover having two dimers and each dimer having two possible spin configurations, we have twelve cartesian states in total. The eight cartesian states of the Kitaev square, shown in Fig.~\ref{fig.sq_cartesian}, are also cartesian states of the tetrahedron. The four additional cartesian states of the tetrahedron are shown in Fig.~\ref{fig.ttet_cartesian}. Note that the Kitaev square CGSS can be viewed as a slice of the tetrahedron CGSS. The tetrahedron CGSS of Fig.~\ref{fig.eightspheres} is embedded in six dimensions spanned by (xyzuvw). Its subset that is contained in the four-dimensional space spanned by (xyuv) gives the Kitaev square CGSS of Fig.~\ref{fig.fourcircles}.

In the context of the Kitaev problem on the honeycomb lattice, BSS pointed out that cartesian states could be connected by smooth energy-preserving transformations. This property holds for the tetrahedron as well. In the geometric picture of the CGSS, these transformations are the great circles along the axis planes.  
In Fig.~\ref{fig.eightspheres}, we see several such smooth transformations that connect cartesian states. For example, $Z$ is connected to $X$, $\bar{X}$, $Y$, $\bar{Y}$, $U$, $\bar{U}$ and $V$, $\bar{V}$ by quarter-arcs. At the same time, we note that there are pairs of cartesian states that are not connected by simple arcs. For example, starting from $Z$, we cannot reach $W$ or $\bar{W}$ via simple arcs. However, we may reach these points by multiple segments, e.g., by going through $X$. This lack of direct connectivity can be understood from the analysis in BSS. The cartesian states $Z$ and $W$ correspond to the same dimer cover, but with different spin orientations on a given dimer. The BSS transformation does not connect such states. 

In the honeycomb lattice Kitaev problem, Chandra et. al. show that the space of ground states is much larger than the set of cartesian states and the valleys that connect them\cite{Chandra2010}. However, it is somewhat difficult to construct these additional states following their formalism. Here, in the example of the Kitaev tetrahedron, we clearly see this physics at play. The cartesian states form a set of zero-dimensional points, with twelve distinct points along the axes. The valleys that connect them are one-dimensional, forming great circles as shown in Fig.~\ref{fig.eightspheres}. However, the CGSS is clearly much larger with the two-dimensional surfaces of the spheres. These additional states lead to new connection pathways on the CGSS.

In summary, the CGSS is composed of eight spheres, embedded in six dimensions. Intersections between spheres make this space a non-manifold. Certain pairs of spheres intersect along great circles while certain pairs only share a common diameter. We also have pairs of spheres that do not intersect at all. 
The nature of the CGSS is much more involved than the intersecting circles of the Kitaev square. In the Kitaev square, the CGSS was generically one-dimensional (circles) with zero-dimensional intersections (points). Here, the CGSS is generically two-dimensional (spheres) with intersections that are one-dimensional (circles) and zero-dimensional (points).

\subsection{Particle on two intersecting sheets}
\label{ssec.twosheets}
Before describing the quantum eigenvalue spectrum of the Kitaev tetrahedron, we discuss a toy problem that gives us a suitable framework. We expect the low energy physics of the Kitaev tetrahedron to map to a single particle problem, where the particle moves on the CGSS space of eight intersecting spheres. This space is a non-manifold that appears to be two-dimensional at a generic point, but has one-dimensional and zero-dimensional self-intersections. What is the low energy behaviour of a particle residing in this space? The insight gleaned from the two circle problem in Sec.~\ref{sec.twocircles} does not suffice to address this question. Working in the same spirit, we construct the simplest toy problem that has the same type of self-intersections. 

\begin{figure}
\includegraphics[width=3in]{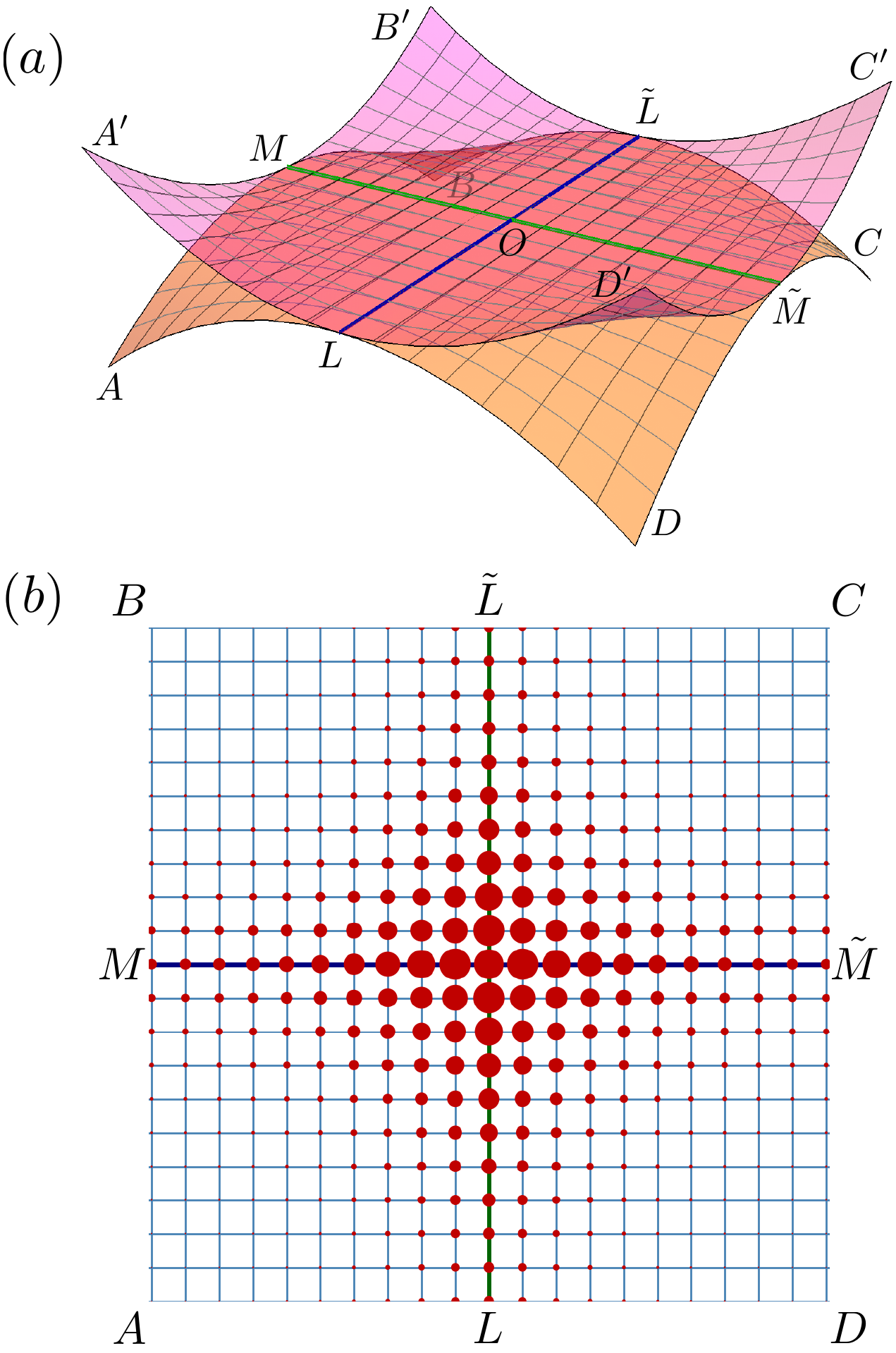}
\caption{Toy problem with two sheets. (a) Geometry of the two sheets intersecting along perpendicular lines. (b) Ground state wavefunction on one of the two sheets. The size of the marker at each site is proportional to the squared amplitude at the site. All sites have the same phase. We show only one sheet as the wavefunction is the same on the other sheet as well. }
\label{fig.twosheets}
\end{figure}

We consider a space composed of two sheets, as shown in Fig~\ref{fig.twosheets}(top). The sheets, $ABCD$ and $A'B'C'D'$ are taken to be squares. Each sheet is taken to have periodic boundaries with opposite sides identified, i.e., $AB \equiv DC$, $AD\equiv BC$,  $A'B' \equiv D'C'$ and $A'D'\equiv B'C'$. The two sheets are assumed to intersect along two perpendicular lines, $L\tilde{L}$ and $M\tilde{M}$, with these lines intersecting at a point $O$. Note that $L$ and $\tilde{L}$ represent the same point due to periodic boundaries, as do $M$ and $\tilde{M}$. %Crucially, the lines $L\tilde{L}$ and $M\tilde{M}$ are common to both sheets. 
This geometry represents the simplest non-manifold space that has the same qualitative features as the Kitaev tetrahedron CGSS. At generic points, it appears two-dimensional. However, it has intersections that are one-dimensional and zero-dimensional. The former are the lines $L\tilde{L}$ and $M\tilde{M}$, while the latter is the point $O$ that lies at the intersection of $L\tilde{L}$ and $M\tilde{M}$.

In order to study the dynamics of a particle on this space, we discretize this space and adopt a tight binding approach. The squares $ABCD$ and $A'B'C'D'$ are both replaced with $N\times N$ meshes with periodic boundaries. Points along the common lines $L\tilde{L}$ and $M\tilde{M}$ are identified. A generic point in this tight binding problem has four nearest neighbours that lie on the same sheet. A point that lies on one of the intersection lines, but not on the other, has six nearest neighbours: two on the common line and two on each sheet. Finally, the common point $O$ has four nearest neighbours: two on each intersection line with all four points shared by both sheets. We construct the corresponding tight binding Hamiltonian and diagonalize it numerically. The resulting spectrum contains, in order of increasing energy, (i) a sharply localized ground state that is centred at $O$ and decays in all directions, (ii) a large number of semi-localized states that are peaked along one of the common lines, (iii) extended states. We focus on the ground state that provides a truly localized state. Its wavefunction is plotted in Fig.~\ref{fig.twosheets}(bottom).  

We now summarize the lessons learned from the toy problem. We have studied a particle moving on a space with two one-dimensional intersection lines. The lines themselves intersect, giving rise to a zero-dimensional singular point. In this scenario, we find only one truly localized state that is centred on the zero-dimensional singularity. Crucially, as the lowest energy state, this state dominates the low energy dynamics of the particle. We now make a connection to the Kitaev tetrahedron CGSS shown in Fig.~\ref{fig.eightspheres}. We have spheres that intersect along lines, with the lines themselves intersecting at points. 

\begin{figure*}
\includegraphics[width=2\columnwidth]{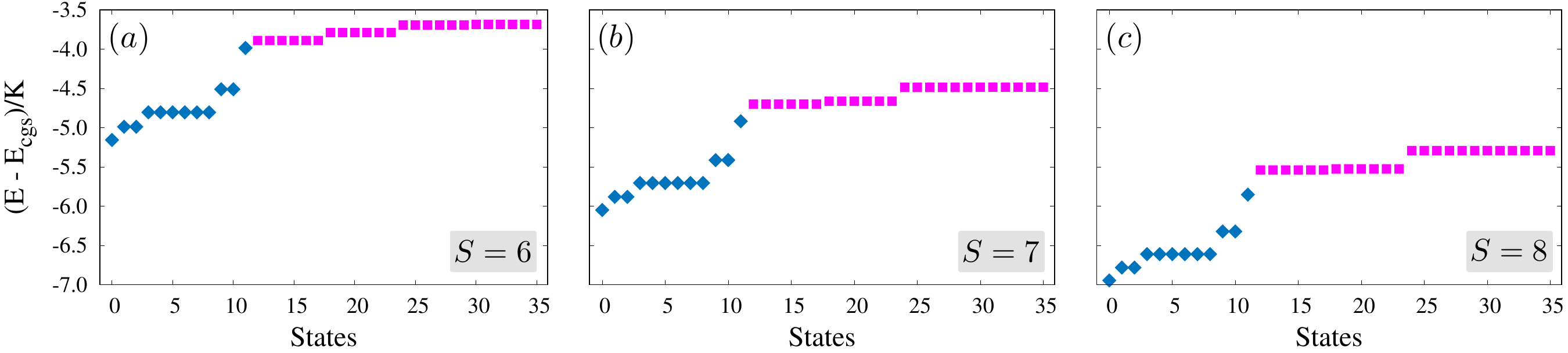}
\caption{Low energy spectra in the Kitaev tetrahedron for $S=6,7,8$ (a, b, c respectively). The spectra are shifted by the classical ground state energy and scaled by the coupling strength $K$. The lowest twelve states are shown using blue diamonds, in order to distinguish them from higher states (magenta squares).}
\label{fig.spectrum_tet}
\end{figure*}

\begin{figure}
\includegraphics[width=2.5in]{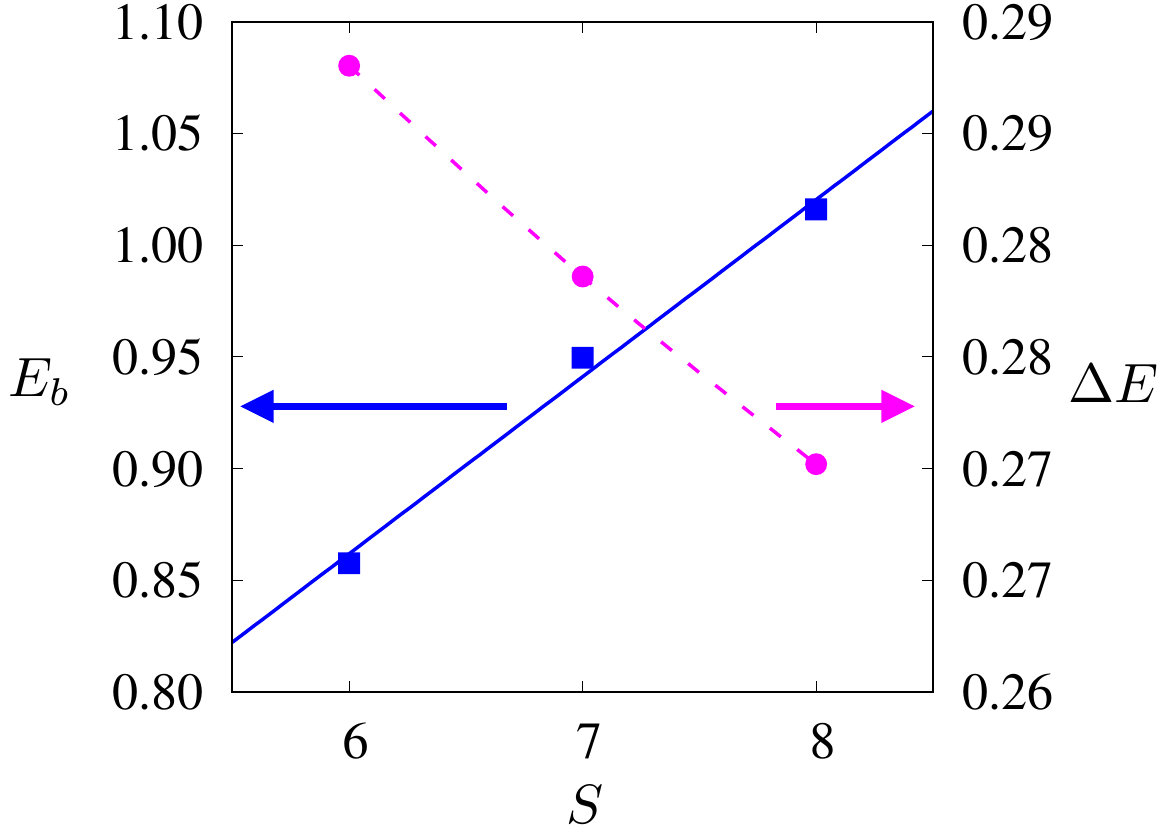}
\caption{Binding energy and spread of the low-lying set of twelve eigenvalues. The binding energy is shown using blue squares. The plotted curve is the fitting function, given by $E_b (S) = 0.386788 + 0.079194 S$. The spread is shown as magenta circles. The dashed line is a guide to the eye.  
}
\label{fig.tet_binding_spread}
\end{figure}

\subsection{Spectrum in the spin-$S$ quantum problem}
The low energy spectrum for the Kitaev tetrahedron is shown in Fig.~\ref{fig.spectrum_tet} for various $S$ values. We interpret its features in terms of the single particle problem on the eight-sphere-CGSS of Fig.~\ref{fig.eightspheres}. Crucially, the lowest energy states are a set of twelve levels. As $S$ increases, the twelve progressively separate from the other, higher energy, states. Fig.~\ref{fig.spectrum_tet} shows the spectrum for $S=6,7,8$. For smaller $S$ values, we find that the twelve states do not separate out completely from the higher states. This is possibly due to the presence of many one-dimensional singularities in the CGSS. In the problem of two intersecting sheets in Sec.~\ref{ssec.twosheets}, there are many `semi-localized' states that are centred on one-dimensional lines. We expect a large number of such states in the eight-sphere CGSS. As they hybridize with one another, they can acquire a large spread in energy. We believe the lowest state from this set is comparable in energy with the highest of the low-lying twelve-fold set. For small $S$ values, this makes it difficult to identify the twelve low-lying states from the numerics. As $S$ increases, this hybridization decreases, with the twelve-fold set becoming clearly visible for $S\geq 5.5$.

We plot two quantities that describe the low energy spectrum in Fig.~\ref{fig.tet_binding_spread}. As in the Kitaev square, we define the binding energy as $E_b = E_{13} - \bar{E}$. Here, $E_{13}$ is the energy of the thirteenth state and $\bar{E}$ is the mean of the twelve lowest energy states. We see that $E_b$ increases linearly in $S$, in line with the scaling arguments in Sec.~\ref{ssec.scaling}. We also plot the spread, $\Delta E$, defined as the standard deviation of the twelve lowest energy states. In analogy with Kitaev square, we expect the spread to decrease with increasing $S$ and to vanish in the $S\rightarrow\infty$ limit. Indeed, we see that the spread decreases with $S$. Due to the limited number of data points, we are not able to find a meaningful fitting function that describes $\Delta E (S)$. As a consequence, we cannot quantitatively address the $S\rightarrow \infty$ limit.
Nevertheless, in analogy with the Kitaev square, we surmise that a twelve-fold degenerate ground state emerges as $S\rightarrow\infty$. We argue that these states are related to the twelve cartesian states of the tetrahedron. This can be understood from the problem of two intersecting sheets in Sec.~\ref{ssec.twosheets} which had a non-degenerate ground state, localized at the point-like singularity. Here, the CGSS of the tetrahedron has twelve point-like singularities, corresponding to cartesian states. The twelve low-lying states arise from bound states around these twelve points. We discuss quantitative tests of this notion below.

\subsection{Character of low lying states }
We next present a test of the hypothesis that the twelve low-lying states are essentially composed of cartesian states. We follow the same steps as in Sec.~\ref{ssec.weight} above to quantify the cartesian weight in the low-lying states. The only difference is that we have 12 Cartesian states on the tetrahedron as opposed to 8 on the square.

\begin{figure}
\includegraphics[width=2.5in]{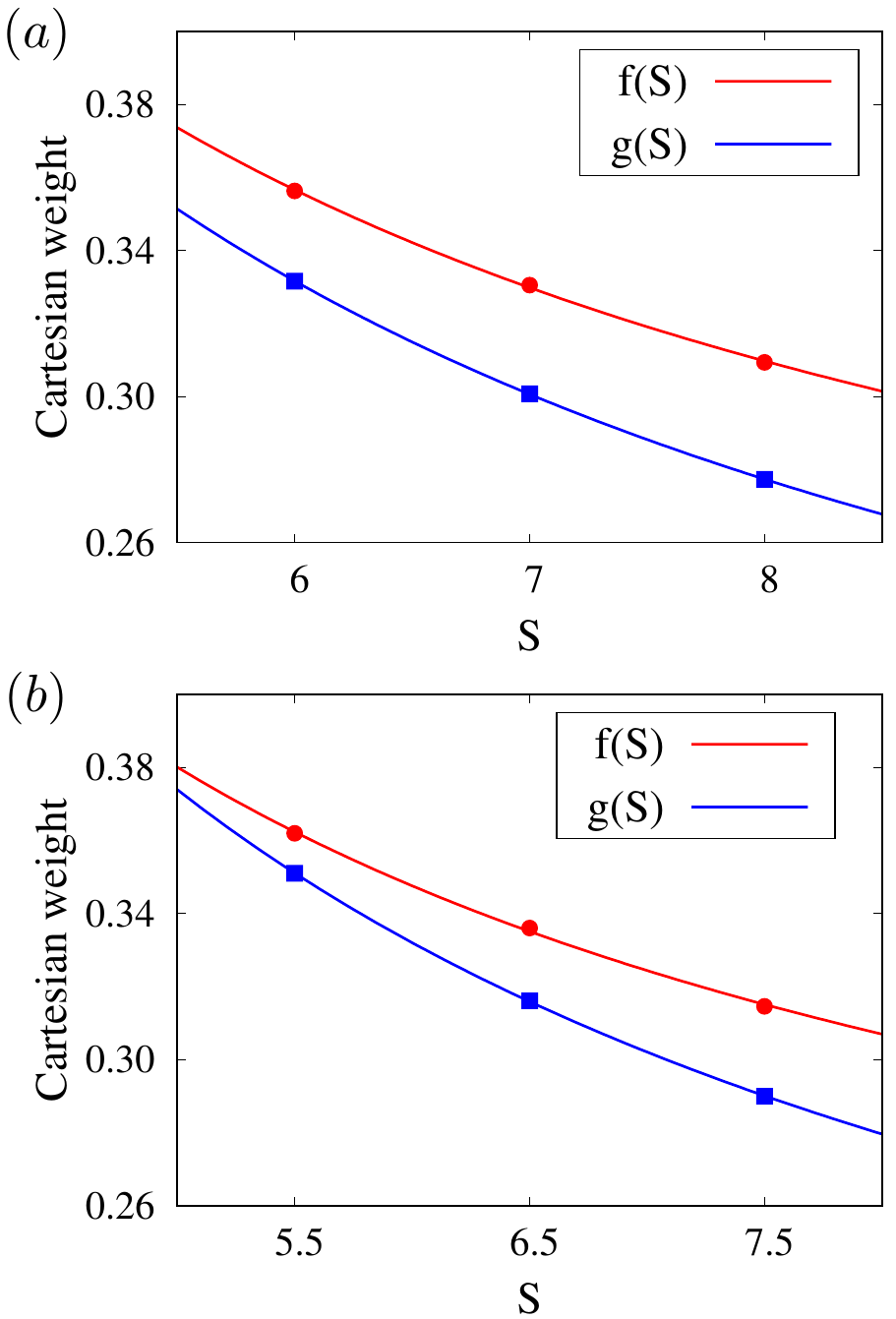}
\caption{Cartesian weight in the low-lying states of the Kitaev tetrahedron for (a) integer and (b) half-integer values of $S$. The blue squares show the cartesian weight in the ground state vs. $S$. The red circles show the average cartesian weight of the twelve low-lying states. For integer $S$ (top), the data are fit using $f(S) = 0.16907 + 1.12538/S $ and $g(S) = 0.11449 + 1.30306/S $. For half-integer $S$ (bottom), the fitting curves are $f(S) = 0.18512 + 0.97489/S $ and $g(S) = 0.12233 + 1.2582/S $.}
\label{fig.tet_proj}
\end{figure}

In Fig.~\ref{fig.tet_proj}, we plot the cartesian weight of the low-lying states of the Kitaev tetrahedron. The two curves correspond to cartesian weight (a) of the ground state (the state with the lowest energy among the twelve), and (b) averaged over the twelve low-lying states. We find sizeable cartesian weight in both. For example, the ground state at $S=7$ has a cartesian weight of $\sim 30\%$. Here, the twelve cartesian states are a minuscule subset of the full Hilbert space that contains $15^4 = 50,625$ states. Nevertheless, they constitute more than a quarter of the weight in the ground state. 

Fig.~\ref{fig.tet_proj} shows the $S$-dependence of the cartesian weights. As with the Kitaev square, we find a smooth variation with $S$ only if we separate integer and half-integer cases. This indicates a role for Berry phases, that is beyond the scope of our discussion.  
Crucially, in both integer and half-integer cases, the cartesian weight extrapolates to non-zero values as $S\rightarrow\infty$. In this limit, the full Hilbert space grows exponentially while the number of cartesian weights remains fixed at twelve. And yet, the cartesian states support a finite fraction of the ground state weight. We interpret this result as follows: the low-lying states are admixtures of bound states formed around zero-dimensional intersections in the CGSS. Their cartesian weight is less than $100\%$ because the bound states are not delta-function-localized. They contain contributions from non-cartesian states that are in the vicinity of the intersection points. 

These results, put together, show that the twelve low-lying states are essentially composed of cartesian states. In this sense, the cartesian states solely determine the low-energy physics of the Kitaev tetrahedron. 

\subsection{Cartesian fidelity of the low-lying states}
\label{ssec.tet_projection}
As with the Kitaev square, we next discuss an independent quantitative test for the cartesian character of the low-lying states. We show that the low-lying states can be mixed into a form that reproduces the twelve classical cartesian states. We define a resolving operator,
\begin{eqnarray}
 \hat{O}_{tet,res.} &=&  \hat{O}_{sq,res.}+ \lambda_{13} (\hat{S}_1^z - \hat{S}_3^z  ) + \lambda_{24}  (\hat{S}_2^z - \hat{S}_4^z  ),~~~~
\end{eqnarray}
where $\hat{O}_{sq,res.}$ has been defined in Eq.~\ref{eq.Osq}. We have two additional terms due to the two additional bonds in the tetrahedron Hamiltonian. As discussed in Sec.~\ref{ssec.fidelity}, these terms resolve cartesian states with dimers on the $z-z$ bonds. 

\begin{figure}
\includegraphics[width=2.5in]{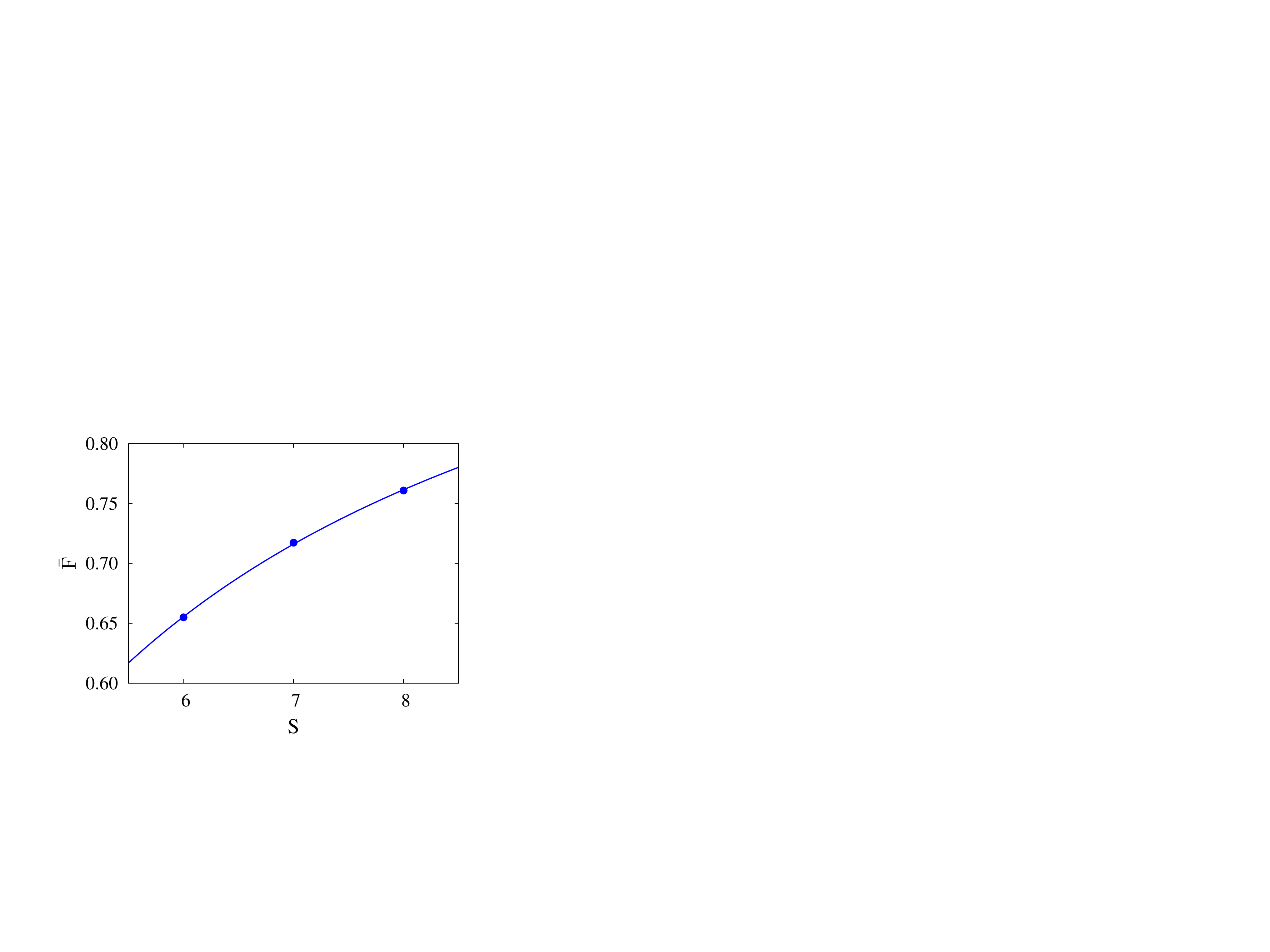}
\caption{ Cartesian fidelity in the Kitaev tetrahedron vs. $S$. The data is fit using $\bar{F}(S) = 1.07978 - 2.54565/S $. }
\label{fig.tet_fidelity}
\end{figure}

Starting the matrix elements of $\hat{O}_{tet,res.}$ in the low-lying states, we define the cartesian fidelity as in Sec.~\ref{ssec.fidelity} above. All details of the definition carry over from the Kitaev square to the tetrahedron, but for the number of cartesian states changing from eight to twelve.
In Fig.~\ref{fig.tet_fidelity}, we plot $\bar{F}$, i.e., $F_\alpha$, averaged over all $\alpha$, for various values of $S$. We see that $\bar{F}$ is very large, increases with $S$ and extrapolates to unity as $S\rightarrow \infty$. This demonstrates that the twelve low-lying states can be mixed to recover cartesian states with high fidelity. 

\section{Summary and discussion}
We have presented low energy descriptions for two Kitaev clusters, the square and the tetrahedron. We have explicitly enumerated the classical ground states in each case, demonstrating that they form self-intersecting spaces. The Kitaev square leads to a space with four circles embedded in four dimensions, while the tetrahedron leads to eight spheres that are embedded in six dimensions. 
We understand the low energy spectra of the spin clusters in terms of a particle moving on these spaces. In both clusters, the low-energy dynamics of this particle is determined by bound states that form at self-intersection points. The intersection points have a very interesting interpretation as `Cartesian' states that were first proposed in Ref.~\onlinecite{Baskaran2008}. Our results show conclusively that Cartesian states, although few in number, completely determine the low energy physics. The validity of this picture improves with $S$, becoming exact in the classical $S\rightarrow\infty$ limit. Our results shed light on the semiclassical physics of Kitaev-type problems. More generally, they provide an enlightening example of order by singularity.

The Kitaev model on the honeycomb lattice also exhibits strong frustration for higher spins. Theoretical studies have discussed possible interesting features for $S\geq 1$\cite{Koga2018,Oitmaa2018,Suzuki2018,Minakawa2019}. An interesting proposal has been put forth for realizing higher-spin Kitaev models in materials\cite{Stavropoulos2019}. In the large-$S$ limit, previous studies have sketched the contours of the ground state space, using the notion of Cartesian states as a convenient starting point. Our results on two clusters suggest a fresh perspectives that prompts a reexamination of earlier results. In particular, on the honeycomb lattice, Chandra et. al.\cite{Chandra2010} have suggested that the space of ground states is a manifold with Cartesian states as extremal points. From an analysis of thermal fluctuations, they further argue that there is no selection of Cartesian states. Here, we have explicitly demonstrated that the two clusters have self-intersecting ground state spaces with clear non-manifold nature. Our analysis also shows that Cartesian states are strongly selected, not by fluctuation contributions to energy but by bound state formation. An exciting future direction is to investigate whether these features carry over to the honeycomb lattice.

There is a large body of work on Kitaev-Heisenberg models, where Kitaev interactions coexist with Heisenberg couplings\cite{Chaloupka2010}. The effects of an additional Heisenberg interaction have also been studied in the classical limit\cite{Price2013}. In this context, our results on the Kitaev-tetrahedron assume significance. We have characterized the space of classical ground states in the pure Kitaev limit. This space of eight spheres persists as the ground state space even in the presence of antiferromagnetic Heisenberg couplings. This can be seen as follows. On the tetrahedron, the Heisenberg interaction can be re-expressed as the square of the total spin. It is minimized in configurations where the sum of four spins vanishes. The set of all such states has been shown to form a non-manifold space that is generically five-dimensional\cite{Khatua2018}. Here, as seen from Tab.~\ref{tab.tetCGSS}, all configurations in the Kitaev-tetrahedron-CGSS have zero total spin. Thus, they continue to be minimum energy states when a Heisenberg interaction is introduced. This indicates that order by singularity operates in the Kitaev-Heisenberg tetrahedron as well.
The Kitaev-tetrahedron-CGSS of Fig.~\ref{fig.eightspheres} can be viewed as a slice of the larger non-manifold CGSS of a Heisenberg tetrahedron.

Our analysis has strong overlaps with the study of quantum graphs. The low energy physics of the Kitaev clusters is dominated by bound states formed at intersection points. These points are connected by pathways, allowing for hybridization among the bound states. At low energies, the clusters can be faithfully modelled as a set of discrete points that are connected by channels. This has strong connections to quantum graph models\cite{Pauling1936,Kottos1997,Keating2008,Harrison2011,Alexandradinata2018}. It is conceivable that more general Kitaev models can be modelled as quantum graphs with a larger number of nodes and connecting pathways. This could allow for new ways of understanding Kitaev spin liquids.

The results presented here are a convincing demonstration of order by singularity. This is only the second known example, after the quantum XY quadrumer studied in Ref.~\onlinecite{Khatua2019}. In Kitaev clusters, Cartesian states are essentially a classical concept. Nevertheless, they acquire a dominant role in the quantum problem. In the $S\rightarrow\infty$ limit, the quantum ground states become the same as cartesian states. They are separated from other classical ground states by an $\mathcal{O}(S)$ binding energy. 
This opens the door to several interesting questions. Is there state selection in the purely classical model with thermal fluctuations? Can order by singularity be distinguished from order by disorder in an experimental context? We hope future studies will be able to answer these questions.

\appendix

\section{Derivation of the CGSS of the Kitaev square}
\label{App.squareCGSS}

We follow the approach of BSS in Ref.~\onlinecite{Baskaran2008} to enumerate classical ground states. As is appropriate for the classical limit, we treat the spins as vectors with three real scalar components. The Kitaev square Hamiltonian is given by 
\begin{equation}
H = K \left[\X{1} \X{2} +\Y{2}\Y{3} + \X{3}\X{4} +\Y{4}\Y{1} \right],
\end{equation}
with $K>0$. We seek to minimize this Hamiltonian over the space of all possible spin configurations. We have twelve variational parameters arising from three components in each spin. However, the minimization is subject to four constraints that fix the spin lengths to be $S$. We use the method of Lagrange multipliers, introducing
\begin{equation}
H_\lambda = \frac{K}{2}\sum_{i=1}^4 \lambda_i \big\{(\X{i})^2+(\Y{i})^2+(\Z{i})^2 - S^2 \big\}.
\end{equation}
The minimization conditions are now given by $\partial {(H - H_\lambda)}/\partial S_i^\alpha = 0$, where $i=1,\ldots,4$ and $\alpha=x,y,z$.
This immediately leads to $\Z{i}=0$ for all $i$, i.e., all four spins lie in the XY plane. In addition, for each bond that connects spins $S_i$ and $S_j$ in the direction $\alpha$ ($x$ or $y$), we obtain 
\begin{equation}
\label{lm1}
S_j^\alpha = \lambda_i S_i^\alpha \textbf{ ; } ~~S_i^\alpha = \lambda_j S_j^\alpha. 
\end{equation}
This represents eight separate equations arising from the four bonds in the problem. Substituting the second equation in the first, we find
\begin{equation}\label{lm2}
S_j^\alpha =( \lambda_i \lambda_j) S_j^\alpha \Rightarrow \lambda_i \lambda_j = 1,
\end{equation}
unless $S_j^\alpha$ is zero. 
For simplicity, we proceed with the analysis assuming that all in-plane components, $S_i^{x/y}$, are non-zero. We will soon show that relaxing this assumption does not lead to any new ground states.
This immediately implies $\lambda_1 = \lambda_3$ and $\lambda_2=\lambda_4$. Using these relations in the Hamiltonian, we obtain
\begin{eqnarray}
E_{min} = 2 KS^2 \lambda_1 = 2 KS^2 \lambda_2 . 
\end{eqnarray}
We conclude that $\lambda_1=\lambda_2$ in the ground states, with all $\lambda$'s being equal. In order to minimize energy while satisfying Eq.~\ref{lm2}, we set $\lambda_i = -1$. The resulting minimum value for the energy is given by $E_{min} = -2KS^2$.

The classical ground states are to be determined from Eq.~\ref{lm1}, taking all $\lambda$'s to be $(-1)$. To systematically enumerate them, we start with the most general form for the first spin, 
\begin{equation}
\mathbf{S}_1 =  S\{ \cos\phi \hat{x} + \sin\phi \hat{y} \}.
\end{equation}
From Eq.~\ref{lm2}, we obtain two other spin components,
\begin{eqnarray}
\X{2} = -S\cos\phi, ~~~\Y{4} = -S \sin\phi. 
\end{eqnarray}
Since we have no constraints on $\mathbf{S}_3$ so far, we consider a general form with a different angle parameter,
\begin{equation}
\mathbf{S}_3 =S \{ \cos\psi \hat{x} +\sin\psi \hat{y} \}. 
\end{equation}
In turn, this fixes two other components with
\begin{eqnarray}
\X{4} = -S\cos\psi,~~~
\Y{2} = -S \sin\psi.
\end{eqnarray}
To preserve the length of $\mathbf{S}_2$ and $\mathbf{S}_4$, we must have 
\begin{equation}
\sin^2\phi = \sin^2\psi.  
\end{equation}
To satisfy this condition, $\psi$ must take one of four possible values: $\phi,~-\phi,~\pi+\phi,~\pi-\phi$.  
Each choice corresponds to one sector in the CGSS, as listed in Tab.~\ref{tab.squareCGSS}. Each sector is a one-parameter family of states, parametrized by $\phi$. All states correspond to the minimum energy, given by $E_{min} = -2KS^2$.

We now show that the same conditions emerge even if some of the spin components are zero. We first list the equations encoded in Eq.~\ref{lm1},
\begin{eqnarray}
 \X{1} &=& \lambda_2 \X{2}, ~~\X{2} ~=~ \lambda_1 \X{1}, \label{eq.12}\\
 \Y{2} &=& \lambda_3 \Y{3}, ~~\Y{3} ~=~ \lambda_2 \Y{2},\label{eq.23} \\
\X{3} &=& \lambda_4 \X{4}, ~~\X{4} ~=~ \lambda_3 \X{3}, \label{eq.34}\\
 \Y{4} &=& \lambda_1 \Y{1}, ~~\Y{1} ~=~ \lambda_4 \Y{4}. \label{eq.41}
\end{eqnarray}
These are necessary conditions (along with $S_i^z=0$ for all $i$) for a minimum energy solution. We now consider the case of $\X{1} =0$, with the first spin taken to point along the Y axis. In order to preserve the length of the first spin, we must have $\Y{1} = \eta_1 S$, where $\eta_1 = \pm 1$. We now use Eq.~\ref{eq.41} to give $\Y{4} = \lambda_1 \Y{1}= \lambda_1 \eta_1 S$. Note that this implies $\lambda_4 \neq 0$, $\lambda_1 \neq 0$ and $\lambda_1 \lambda_4 =1$. We must also have
$\vert \lambda_1 \vert \leq 1$, as any component of $S_4$ cannot exceed $S$ without violating the spin length constraint. As the $y$ component of $S_4$ is now fixed, we fix its $x$ component to satisfy the spin length condition. We take $\X{4} = \eta_4 \sqrt{1-\lambda_1^2} S$, with $\eta_4 = \pm1$. 

In the same way, we now fix $S_2$ and $S_3$. From Eq.~\ref{eq.12}, we have $\X{2} = 0$ and therefore $\Y{2} = \eta_2 S$, where $\eta_2 = \pm1$. Eq.~\ref{eq.23} gives $\Y{3} = \lambda_2 \Y{2} = \lambda_2 \eta_2 S$. We have $\lambda_2 \neq 0$, $\lambda_3 \neq 0$ and $\lambda_2 \lambda_3 = 1$. We also have $\vert \lambda_2 \vert \leq 1$ as no component can exceed $S$. With $\Y{3}$ known, we fix its $x$ component from the spin length condition, $\X{3} = \eta_3 \sqrt{1-\lambda_2^2} S$, with $\eta_3 =\pm 1$.

We have a constraint in Eq.~\ref{eq.34} that relates $\X{3}$ and $\X{4}$. Using the relation $\X{3} = \lambda_4 \X{4} $, we obtain
\begin{eqnarray}
\eta_3 \sqrt{1-\lambda_2^2} = \eta_4 \lambda_4 \sqrt{1-\lambda_1^2} = \eta_4  \sqrt{\lambda_4^2-1} .
\end{eqnarray}
This gives $\eta_3 = \eta_4$ as well as $\lambda_2^2 + \lambda_4^2 = 2$.

Gathering the four spins together, we have
\begin{eqnarray}
\mathbf{S}_1&=& S(0,\eta_1,0 ), \\
\mathbf{S}_2 &=& S(0,\eta_2,0 ), \\
\mathbf{S}_3 &=& S (\eta_3 \sqrt{1-\lambda_2^2} ,\eta_2 \lambda_2,0), \\
\mathbf{S}_4 &=& S (\eta_4 \sqrt{1-\lambda_1^2} ,\eta_1 \lambda_1,0).
\end{eqnarray}
Substituting these expressions in the Hamiltonian, we have
\begin{eqnarray}
H = KS^2 \big[  \lambda_2 + \sqrt{1-\lambda_1^2}\sqrt{1-\lambda_2^2} + \lambda_1 \big].
\end{eqnarray}
The energy is determined by the $\lambda$ coefficients that must be chosen subject to the constraints, $\vert \lambda_1 \vert \leq 1$ and $\vert \lambda_2 \vert \leq 1$. Clearly, the choice that gives the lowest energy value is $\lambda_1 = \lambda_2 = -1$: the same conclusion that was drawn earlier by assuming that all spin components were non-zero. Here, in fact, our choice of one component being zero fixes the spin configuration to be a Cartesian state. The resulting energy is the same as $E_{min}$ given above. These arguments can be extended to states where any of the other components are zero. This invariably leads to the same minimum energy and the same conditions.

\section{Berry phase effects in the Kitaev square}
\label{App.Berry}
The low energy physics of the Kitaev square maps to a particle moving on the CGSS of four intersecting circles. This mapping is based on the spin path integral formulation, wherein deviations from the CGSS are exponentially suppressed due to their energy cost. Apart from an energy term, the spin path integral has a geometric (Berry) phase term that attaches complex phases to closed paths. In the mapping to the single particle problem, these manifest as Aharonov-Bohm phases arising from flux lines that thread the CGSS. This has been demonstrated in Ref.~\onlinecite{Khatua2019} in the case of the trimer with XY antiferromagnetic couplings. 

We now demonstrate the existence of paths within the Kitaev-square-CGSS that accrue a non-trivial Berry phase. We work in the spin path integral language, where the Berry phase has a simple geometric interpretation, given by
\begin{eqnarray}
S_B = S \sum_{i=1}^4 \mathcal{A}_i,
\end{eqnarray}
where $\mathcal{A}_i$ is the solid angle subtended at the north pole by the spin $\mathbf{S}_i$ as it describes a closed trajectory.

In the CGSS shown in Fig.~\ref{fig.fourcircles}, we consider the path $X \rightarrow Y \rightarrow \bar{U} \rightarrow \bar{V}  \rightarrow X$. This path consists of four arcs, each within one of the circles of the CGSS. In each arc, all four spins move along a quarter-arc in spin space. 
As we move along the entire path, each spin describes a closed trajectory given by
\begin{eqnarray}
\mathbf{S}_1&:&  \hat{x} \rightarrow \hat{y} \rightarrow -\hat{x} \rightarrow -\hat{y} \rightarrow  \hat{x}, \\ 
\mathbf{S}_2&:& -\hat{x} \rightarrow -\hat{y} \rightarrow \hat{x} \rightarrow -\hat{y} \rightarrow  -\hat{x}, \\ 
\mathbf{S}_3&:&  \hat{x} \rightarrow \hat{y} \rightarrow \hat{x} \rightarrow \hat{y} \rightarrow  \hat{x}, \\ 
\mathbf{S}_4&:& -\hat{x} \rightarrow -\hat{y} \rightarrow -\hat{x} \rightarrow \hat{y} \rightarrow  -\hat{x}.
\end{eqnarray}
Here, $\pm \hat{x}$ and $\pm \hat{y}$ represent directions in spin space. Notably, $\mathbf{S}_1$ describes a complete circle in the counter-clockwise direction (looking down from the positive $Z$ axis), whereas the other three spins do not. In other words, the first spin subtends a solid angle of $2\pi$ at the north pole. The other three spins subtend a net-zero solid angle. 

Thus, the Berry phase associate with the path $X \rightarrow Y \rightarrow \bar{U} \rightarrow \bar{V}  \rightarrow X$ is $2\pi S$. For integer $S$ values, this Berry phase is trivial. However, for half-integer spins, this is a physically relevant Berry phase of $\pi$. This explains the qualitative difference in behaviour between half-integer and integer spins seen in Fig.~\ref{fig.sq_proj}. We argue that there is no further distinction beyond integer and half-integer cases, e.g., there is no qualitative difference between even or odd integer values of $S$. This can be seen by the following argument. As all the states in the CGSS are coplanar, each spin is constrained to move along a circle in the XY plane. A closed path traversed by a spin must necessarily be composed of an integral number of circles. Thus, the solid angle subtended at the north pole is an integer multiple of $2\pi$ for each spin. The overall Berry phase for any closed path has the form $2n\pi S$, where $n$ is an integer. This only allows for a distinction between integer and half-integer $S$ values.

We believe there is a role for Berry phases in the Kitaev tetrahedron as well. We see a clear distinction between half-integer and integer $S$ values in Fig.~\ref{fig.tet_proj}. The non-trivial path discussed in the context of the Kitaev square ( $X \rightarrow Y \rightarrow \bar{U} \rightarrow \bar{V}  \rightarrow X$) is a valid path in the tetrahedron-CGSS as well. This shows that Berry phase has a role to play in the tetrahedron. However, unlike the square, we do not have a precise argument as to why the Berry phase can only distinguish half-integer and integer $S$ values. 
As the tetrahedron CGSS is more complex than that of the square, we content ourselves with the empirical observation that separating half-integer and integer values of $S$ leads to smooth $S$-dependence. 

\section{Energy minimization in the Kitaev tetrahedron}
\label{App.tetCGSS}
The classical ground states of the Kitaev tetrahedron can be found in the same manner as for the Kitaev square in Appendix.~\ref{App.squareCGSS} above. This leads to the equations in Eqs.~\ref{eq.12}-\ref{eq.41}. In addition, we have four more equations given by
\begin{eqnarray}
 \Z{1} &=& \lambda_3 \Z{3}, ~~\Z{3} ~=~ \lambda_1 \Z{1}, \label{eq.13}\\
 \Z{2} &=& \lambda_4 \Z{4}, ~~\Z{4} ~=~ \lambda_2 \Z{2},\label{eq.24}.
 \end{eqnarray}
We first make a simplifying assumption, taking all twelve spin components to be non-zero. This leads to $\lambda_i \lambda_j = 1$, for all pairs $(i,j)$. This implies that all $\lambda$'s are equal, given by $\pm 1$. The classical energy comes out to be $E = 2KS^2 \lambda$. In order to minimize the energy, we choose $\lambda =- 1$. Thus, the ground states are to be chosen from Eqs.~\ref{eq.12}-\ref{eq.41} as well as Eqs.~\ref{eq.13} and \ref{eq.24}, with all $\lambda$'s set to $(-1)$. 

If we allow for some spin components to be zero, we do not obtain any new conditions or ground states. We simply recover a subset of the states obtained from the considerations given above. 
In all cases, the ground state energy is the same as that for the Kitaev square, $E_{min} = -2KS^2$.

\section{Classical ground state in the Kitaev tetrahedron}
\label{App.tetCGSS_b}

To derive the CGSS of the tetrahedron, we start with the most general form for $\mathbf{S}_1$,

\begin{equation}
\mathbf{S}_1 = S \{ \sin{\theta} \cos\phi\hat{x} + \sin{\theta}\sin{\phi}\hat{y}+\cos{\theta}\hat{z}\},
\end{equation}
where $\theta \in [0,\pi]$ and $\phi\in [0,2\pi)$ are arbitrary. From Eqs.~\ref{eq.12}-\ref{eq.41} and \ref{eq.13}-\ref{eq.24}, we obtain
\begin{equation}
\nonumber \X{2} = -S\sin\theta \cos\phi,~~
\Y{4} = -S\sin\theta \sin\phi, ~~
\Z{3} = -S\cos\theta.
\end{equation}
The equation for $\Z{3}$ implies that $(\X{3})^2+(\Y{3})^2=S^2 \sin^2\theta$. Introducing a new angle $\psi$, we have
\begin{equation}
\mathbf{S}_3 = S\{ \sin\theta\cos\psi \hat{x} + \sin\theta\sin\psi\hat{y} -\cos\theta\hat{z}\}.
\end{equation}

This, in turn, fixes the following components:

\begin{equation}
\nonumber \X{4} = -S\sin\theta\cos\psi, ~~~
\Y{2} = -S\sin\theta\sin\psi.
\end{equation}

The only undetermined components now are $\Z{2}$ and $\Z{4}$ which are related by $\Z{2}=-\Z{4}$. This implies $(\X{2})^2+(\Y{2})^2 = (\X{4})^2+(\Y{4})^2$, giving
\begin{equation}
\nonumber \sin^2\theta (\cos^2\psi + \sin^2\phi) = \sin^2\theta (\cos^2\phi +\sin^2\psi).
\end{equation}
This has two solutions. Either $\sin\theta=0$ (which gives just the Cartesian states in the z-direction) or $\sin^2\psi=\sin^2\phi$. The latter has four solutions (as discussed in the Kitaev square): $\psi = \phi,~-\phi,~\pi+\phi,~\pi-\phi$. In all four cases, we find $(\Z{2})^2 = (\Z{4})^2 = S^2 \cos^2\theta$. As $\Z{2}=-\Z{4}$, we have

\begin{equation}
\Z{2} = \pm S \cos\theta, \Z{4} = \mp S \cos\theta.
\end{equation}

We note that there are 8 distinct cases here: four choices for $\psi$ and two possible signs in $\Z{2}$. These choices correspond to the eight sectors of the CGSS described in the main text. In each sector, we have a family of states that is parametrized by two angles, $\theta$ and $\phi$.

In all eight cases, upon substituting the expressions for spin components in the Hamiltonian, we obtain $E =  -2KS^2 = E_{min}$. This confirms that every state in each of the eight sectors is indeed a classical ground state.
\bibliographystyle{apsrev4-1} 
\bibliography{Kitaev_ObS}

%merlin.mbs apsrev4-1.bst 2010-07-25 4.21a (PWD, AO, DPC) hacked
%Control: key (0)
%Control: author (72) initials jnrlst
%Control: editor formatted (1) identically to author
%Control: production of article title (-1) disabled
%Control: page (0) single
%Control: year (1) truncated
%Control: production of eprint (0) enabled
\begin{thebibliography}{39}%
\makeatletter
\providecommand \@ifxundefined [1]{%
 \@ifx{#1\undefined}
}%
\providecommand \@ifnum [1]{%
 \ifnum #1\expandafter \@firstoftwo
 \else \expandafter \@secondoftwo
 \fi
}%
\providecommand \@ifx [1]{%
 \ifx #1\expandafter \@firstoftwo
 \else \expandafter \@secondoftwo
 \fi
}%
\providecommand \natexlab [1]{#1}%
\providecommand \enquote  [1]{``#1''}%
\providecommand \bibnamefont  [1]{#1}%
\providecommand \bibfnamefont [1]{#1}%
\providecommand \citenamefont [1]{#1}%
\providecommand \href@noop [0]{\@secondoftwo}%
\providecommand \href [0]{\begingroup \@sanitize@url \@href}%
\providecommand \@href[1]{\@@startlink{#1}\@@href}%
\providecommand \@@href[1]{\endgroup#1\@@endlink}%
\providecommand \@sanitize@url [0]{\catcode `\\12\catcode `\$12\catcode
  `\&12\catcode `\#12\catcode `\^12\catcode `\_12\catcode `\%12\relax}%
\providecommand \@@startlink[1]{}%
\providecommand \@@endlink[0]{}%
\providecommand \url  [0]{\begingroup\@sanitize@url \@url }%
\providecommand \@url [1]{\endgroup\@href {#1}{\urlprefix }}%
\providecommand \urlprefix  [0]{URL }%
\providecommand \Eprint [0]{\href }%
\providecommand \doibase [0]{http://dx.doi.org/}%
\providecommand \selectlanguage [0]{\@gobble}%
\providecommand \bibinfo  [0]{\@secondoftwo}%
\providecommand \bibfield  [0]{\@secondoftwo}%
\providecommand \translation [1]{[#1]}%
\providecommand \BibitemOpen [0]{}%
\providecommand \bibitemStop [0]{}%
\providecommand \bibitemNoStop [0]{.\EOS\space}%
\providecommand \EOS [0]{\spacefactor3000\relax}%
\providecommand \BibitemShut  [1]{\csname bibitem#1\endcsname}%
\let\auto@bib@innerbib\@empty
%</preamble>
\bibitem [{\citenamefont {Chalker}(2011)}]{Chalker2011}%
  \BibitemOpen
  \bibfield  {author} {\bibinfo {author} {\bibfnamefont {J.~T.}\ \bibnamefont
  {Chalker}},\ }\enquote {\bibinfo {title} {Geometrically frustrated
  antiferromagnets: Statistical mechanics and dynamics},}\ in\ \href {\doibase
  10.1007/978-3-642-10589-0_1} {\emph {\bibinfo {booktitle} {Introduction to
  Frustrated Magnetism: Materials, Experiments, Theory}}},\ \bibinfo {editor}
  {edited by\ \bibinfo {editor} {\bibfnamefont {C.}~\bibnamefont {Lacroix}},
  \bibinfo {editor} {\bibfnamefont {P.}~\bibnamefont {Mendels}}, \ and\
  \bibinfo {editor} {\bibfnamefont {F.}~\bibnamefont {Mila}}}\ (\bibinfo
  {publisher} {Springer Berlin Heidelberg},\ \bibinfo {address} {Berlin,
  Heidelberg},\ \bibinfo {year} {2011})\ pp.\ \bibinfo {pages}
  {3--22}\BibitemShut {NoStop}%
\bibitem [{\citenamefont {{Villain, J.}}\ \emph {et~al.}(1980)\citenamefont
  {{Villain, J.}}, \citenamefont {{Bidaux, R.}}, \citenamefont {{Carton,
  J.-P.}},\ and\ \citenamefont {{Conte, R.}}}]{Villain1980}%
  \BibitemOpen
  \bibfield  {author} {\bibinfo {author} {\bibnamefont {{Villain, J.}}},
  \bibinfo {author} {\bibnamefont {{Bidaux, R.}}}, \bibinfo {author}
  {\bibnamefont {{Carton, J.-P.}}}, \ and\ \bibinfo {author} {\bibnamefont
  {{Conte, R.}}},\ }\href {\doibase 10.1051/jphys:0198000410110126300}
  {\bibfield  {journal} {\bibinfo  {journal} {J. Phys. France}\ }\textbf
  {\bibinfo {volume} {41}},\ \bibinfo {pages} {1263} (\bibinfo {year}
  {1980})}\BibitemShut {NoStop}%
\bibitem [{\citenamefont {Shender}(1982)}]{Shender1982}%
  \BibitemOpen
  \bibfield  {author} {\bibinfo {author} {\bibfnamefont {E.~F.}\ \bibnamefont
  {Shender}},\ }\href
  {http://www.jetp.ac.ru/cgi-bin/e/index/e/56/1/p178?a=list} {\bibfield
  {journal} {\bibinfo  {journal} {JETP}\ }\textbf {\bibinfo {volume} {56}},\
  \bibinfo {pages} {178} (\bibinfo {year} {1982})}\BibitemShut {NoStop}%
\bibitem [{\citenamefont {Henley}(1989)}]{Henley1989}%
  \BibitemOpen
  \bibfield  {author} {\bibinfo {author} {\bibfnamefont {C.~L.}\ \bibnamefont
  {Henley}},\ }\href {\doibase 10.1103/PhysRevLett.62.2056} {\bibfield
  {journal} {\bibinfo  {journal} {Phys. Rev. Lett.}\ }\textbf {\bibinfo
  {volume} {62}},\ \bibinfo {pages} {2056} (\bibinfo {year}
  {1989})}\BibitemShut {NoStop}%
\bibitem [{\citenamefont {Khatua}\ \emph {et~al.}(2019)\citenamefont {Khatua},
  \citenamefont {Sen},\ and\ \citenamefont {Ganesh}}]{Khatua2019}%
  \BibitemOpen
  \bibfield  {author} {\bibinfo {author} {\bibfnamefont {S.}~\bibnamefont
  {Khatua}}, \bibinfo {author} {\bibfnamefont {D.}~\bibnamefont {Sen}}, \ and\
  \bibinfo {author} {\bibfnamefont {R.}~\bibnamefont {Ganesh}},\ }\href
  {\doibase 10.1103/PhysRevB.100.134411} {\bibfield  {journal} {\bibinfo
  {journal} {Phys. Rev. B}\ }\textbf {\bibinfo {volume} {100}},\ \bibinfo
  {pages} {134411} (\bibinfo {year} {2019})}\BibitemShut {NoStop}%
\bibitem [{\citenamefont {Kitaev}(2006)}]{Kitaev2006}%
  \BibitemOpen
  \bibfield  {author} {\bibinfo {author} {\bibfnamefont {A.}~\bibnamefont
  {Kitaev}},\ }\href {\doibase https://doi.org/10.1016/j.aop.2005.10.005}
  {\bibfield  {journal} {\bibinfo  {journal} {Annals of Physics}\ }\textbf
  {\bibinfo {volume} {321}},\ \bibinfo {pages} {2 } (\bibinfo {year} {2006})},\
  \bibinfo {note} {january Special Issue}\BibitemShut {NoStop}%
\bibitem [{\citenamefont {Kitaev}\ and\ \citenamefont
  {Laumann}(2009)}]{Kitaev2009}%
  \BibitemOpen
  \bibfield  {author} {\bibinfo {author} {\bibfnamefont {A.}~\bibnamefont
  {Kitaev}}\ and\ \bibinfo {author} {\bibfnamefont {C.}~\bibnamefont
  {Laumann}},\ }\href@noop {} {\enquote {\bibinfo {title} {Topological phases
  and quantum computation},}\ } (\bibinfo {year} {2009}),\ \Eprint
  {http://arxiv.org/abs/0904.2771} {arXiv:0904.2771 [cond-mat.mes-hall]}
  \BibitemShut {NoStop}%
\bibitem [{\citenamefont {Nussinov}\ and\ \citenamefont {van~den
  Brink}(2013)}]{Nussinov2013}%
  \BibitemOpen
  \bibfield  {author} {\bibinfo {author} {\bibfnamefont {Z.}~\bibnamefont
  {Nussinov}}\ and\ \bibinfo {author} {\bibfnamefont {J.}~\bibnamefont {van~den
  Brink}},\ }\href@noop {} {\enquote {\bibinfo {title} {Compass and kitaev
  models -- theory and physical motivations},}\ } (\bibinfo {year} {2013}),\
  \Eprint {http://arxiv.org/abs/1303.5922} {arXiv:1303.5922 [cond-mat.str-el]}
  \BibitemShut {NoStop}%
\bibitem [{\citenamefont {Perreault}(2016)}]{Perreault2016}%
  \BibitemOpen
  \bibfield  {author} {\bibinfo {author} {\bibfnamefont {B.}~\bibnamefont
  {Perreault}},\ }\emph {\bibinfo {title} {Identifying a Kitaev Spin Liquid}},\
  \href {http://hdl.handle.net/11299/185143} {Ph.D. thesis},\ \bibinfo
  {school} {University of Minnesota} (\bibinfo {year} {2016})\BibitemShut
  {NoStop}%
\bibitem [{\citenamefont {Zhou}\ \emph {et~al.}(2017)\citenamefont {Zhou},
  \citenamefont {Kanoda},\ and\ \citenamefont {Ng}}]{Zhou2017}%
  \BibitemOpen
  \bibfield  {author} {\bibinfo {author} {\bibfnamefont {Y.}~\bibnamefont
  {Zhou}}, \bibinfo {author} {\bibfnamefont {K.}~\bibnamefont {Kanoda}}, \ and\
  \bibinfo {author} {\bibfnamefont {T.-K.}\ \bibnamefont {Ng}},\ }\href
  {\doibase 10.1103/RevModPhys.89.025003} {\bibfield  {journal} {\bibinfo
  {journal} {Rev. Mod. Phys.}\ }\textbf {\bibinfo {volume} {89}},\ \bibinfo
  {pages} {025003} (\bibinfo {year} {2017})}\BibitemShut {NoStop}%
\bibitem [{\citenamefont {Rao}(2017)}]{Rao2017}%
  \BibitemOpen
  \bibfield  {author} {\bibinfo {author} {\bibfnamefont {S.}~\bibnamefont
  {Rao}},\ }\enquote {\bibinfo {title} {Introduction to abelian and non-abelian
  anyons},}\ in\ \href {\doibase 10.1007/978-981-10-6841-6_16} {\emph {\bibinfo
  {booktitle} {Topology and Condensed Matter Physics}}},\ \bibinfo {editor}
  {edited by\ \bibinfo {editor} {\bibfnamefont {S.~M.}\ \bibnamefont
  {Bhattacharjee}}, \bibinfo {editor} {\bibfnamefont {M.}~\bibnamefont {Mj}}, \
  and\ \bibinfo {editor} {\bibfnamefont {A.}~\bibnamefont {Bandyopadhyay}}}\
  (\bibinfo  {publisher} {Springer Singapore},\ \bibinfo {address}
  {Singapore},\ \bibinfo {year} {2017})\ pp.\ \bibinfo {pages}
  {399--437}\BibitemShut {NoStop}%
\bibitem [{\citenamefont {Hermanns}\ \emph {et~al.}(2018)\citenamefont
  {Hermanns}, \citenamefont {Kimchi},\ and\ \citenamefont
  {Knolle}}]{Hermanns2018}%
  \BibitemOpen
  \bibfield  {author} {\bibinfo {author} {\bibfnamefont {M.}~\bibnamefont
  {Hermanns}}, \bibinfo {author} {\bibfnamefont {I.}~\bibnamefont {Kimchi}}, \
  and\ \bibinfo {author} {\bibfnamefont {J.}~\bibnamefont {Knolle}},\ }\href
  {\doibase 10.1146/annurev-conmatphys-033117-053934} {\bibfield  {journal}
  {\bibinfo  {journal} {Annual Review of Condensed Matter Physics}\ }\textbf
  {\bibinfo {volume} {9}},\ \bibinfo {pages} {17} (\bibinfo {year} {2018})},\
  \Eprint
  {http://arxiv.org/abs/https://doi.org/10.1146/annurev-conmatphys-033117-053934}
  {https://doi.org/10.1146/annurev-conmatphys-033117-053934} \BibitemShut
  {NoStop}%
\bibitem [{\citenamefont {Rau}\ \emph {et~al.}(2016)\citenamefont {Rau},
  \citenamefont {Lee},\ and\ \citenamefont {Kee}}]{Rau2016}%
  \BibitemOpen
  \bibfield  {author} {\bibinfo {author} {\bibfnamefont {J.~G.}\ \bibnamefont
  {Rau}}, \bibinfo {author} {\bibfnamefont {E.~K.-H.}\ \bibnamefont {Lee}}, \
  and\ \bibinfo {author} {\bibfnamefont {H.-Y.}\ \bibnamefont {Kee}},\ }\href
  {\doibase 10.1146/annurev-conmatphys-031115-011319} {\bibfield  {journal}
  {\bibinfo  {journal} {Annual Review of Condensed Matter Physics}\ }\textbf
  {\bibinfo {volume} {7}},\ \bibinfo {pages} {195} (\bibinfo {year} {2016})},\
  \Eprint
  {http://arxiv.org/abs/https://doi.org/10.1146/annurev-conmatphys-031115-011319}
  {https://doi.org/10.1146/annurev-conmatphys-031115-011319} \BibitemShut
  {NoStop}%
\bibitem [{\citenamefont {Takagi}\ \emph {et~al.}(2019)\citenamefont {Takagi},
  \citenamefont {Takayama}, \citenamefont {Jackeli}, \citenamefont
  {Khaliullin},\ and\ \citenamefont {Nagler}}]{Takagi2019}%
  \BibitemOpen
  \bibfield  {author} {\bibinfo {author} {\bibfnamefont {H.}~\bibnamefont
  {Takagi}}, \bibinfo {author} {\bibfnamefont {T.}~\bibnamefont {Takayama}},
  \bibinfo {author} {\bibfnamefont {G.}~\bibnamefont {Jackeli}}, \bibinfo
  {author} {\bibfnamefont {G.}~\bibnamefont {Khaliullin}}, \ and\ \bibinfo
  {author} {\bibfnamefont {S.~E.}\ \bibnamefont {Nagler}},\ }\href {\doibase
  10.1038/s42254-019-0038-2} {\bibfield  {journal} {\bibinfo  {journal} {Nature
  Reviews Physics}\ }\textbf {\bibinfo {volume} {1}},\ \bibinfo {pages} {264}
  (\bibinfo {year} {2019})}\BibitemShut {NoStop}%
\bibitem [{\citenamefont {Baskaran}\ \emph {et~al.}(2008)\citenamefont
  {Baskaran}, \citenamefont {Sen},\ and\ \citenamefont
  {Shankar}}]{Baskaran2008}%
  \BibitemOpen
  \bibfield  {author} {\bibinfo {author} {\bibfnamefont {G.}~\bibnamefont
  {Baskaran}}, \bibinfo {author} {\bibfnamefont {D.}~\bibnamefont {Sen}}, \
  and\ \bibinfo {author} {\bibfnamefont {R.}~\bibnamefont {Shankar}},\ }\href
  {\doibase 10.1103/PhysRevB.78.115116} {\bibfield  {journal} {\bibinfo
  {journal} {Phys. Rev. B}\ }\textbf {\bibinfo {volume} {78}},\ \bibinfo
  {pages} {115116} (\bibinfo {year} {2008})}\BibitemShut {NoStop}%
\bibitem [{\citenamefont {Chandra}\ \emph {et~al.}(2010)\citenamefont
  {Chandra}, \citenamefont {Ramola},\ and\ \citenamefont {Dhar}}]{Chandra2010}%
  \BibitemOpen
  \bibfield  {author} {\bibinfo {author} {\bibfnamefont {S.}~\bibnamefont
  {Chandra}}, \bibinfo {author} {\bibfnamefont {K.}~\bibnamefont {Ramola}}, \
  and\ \bibinfo {author} {\bibfnamefont {D.}~\bibnamefont {Dhar}},\ }\href
  {\doibase 10.1103/PhysRevE.82.031113} {\bibfield  {journal} {\bibinfo
  {journal} {Phys. Rev. E}\ }\textbf {\bibinfo {volume} {82}},\ \bibinfo
  {pages} {031113} (\bibinfo {year} {2010})}\BibitemShut {NoStop}%
\bibitem [{\citenamefont {Rousochatzakis}\ \emph {et~al.}(2018)\citenamefont
  {Rousochatzakis}, \citenamefont {Sizyuk},\ and\ \citenamefont
  {Perkins}}]{Rousochatzakis2018}%
  \BibitemOpen
  \bibfield  {author} {\bibinfo {author} {\bibfnamefont {I.}~\bibnamefont
  {Rousochatzakis}}, \bibinfo {author} {\bibfnamefont {Y.}~\bibnamefont
  {Sizyuk}}, \ and\ \bibinfo {author} {\bibfnamefont {N.~B.}\ \bibnamefont
  {Perkins}},\ }\href {\doibase 10.1038/s41467-018-03934-1} {\bibfield
  {journal} {\bibinfo  {journal} {Nature Communications}\ }\textbf {\bibinfo
  {volume} {9}},\ \bibinfo {pages} {1575} (\bibinfo {year} {2018})}\BibitemShut
  {NoStop}%
\bibitem [{\citenamefont {Saket}\ \emph {et~al.}(2010)\citenamefont {Saket},
  \citenamefont {Hassan},\ and\ \citenamefont {Shankar}}]{Saket2010}%
  \BibitemOpen
  \bibfield  {author} {\bibinfo {author} {\bibfnamefont {A.}~\bibnamefont
  {Saket}}, \bibinfo {author} {\bibfnamefont {S.~R.}\ \bibnamefont {Hassan}}, \
  and\ \bibinfo {author} {\bibfnamefont {R.}~\bibnamefont {Shankar}},\ }\href
  {\doibase 10.1103/PhysRevB.82.174409} {\bibfield  {journal} {\bibinfo
  {journal} {Phys. Rev. B}\ }\textbf {\bibinfo {volume} {82}},\ \bibinfo
  {pages} {174409} (\bibinfo {year} {2010})}\BibitemShut {NoStop}%
\bibitem [{\citenamefont {Mandal}\ and\ \citenamefont
  {Surendran}(2009)}]{Mandal2009}%
  \BibitemOpen
  \bibfield  {author} {\bibinfo {author} {\bibfnamefont {S.}~\bibnamefont
  {Mandal}}\ and\ \bibinfo {author} {\bibfnamefont {N.}~\bibnamefont
  {Surendran}},\ }\href {\doibase 10.1103/PhysRevB.79.024426} {\bibfield
  {journal} {\bibinfo  {journal} {Phys. Rev. B}\ }\textbf {\bibinfo {volume}
  {79}},\ \bibinfo {pages} {024426} (\bibinfo {year} {2009})}\BibitemShut
  {NoStop}%
\bibitem [{\citenamefont {Trebst}(2017)}]{Trebst2017}%
  \BibitemOpen
  \bibfield  {author} {\bibinfo {author} {\bibfnamefont {S.}~\bibnamefont
  {Trebst}},\ }\href@noop {} {\enquote {\bibinfo {title} {Kitaev materials},}\
  } (\bibinfo {year} {2017}),\ \Eprint {http://arxiv.org/abs/1701.07056}
  {arXiv:1701.07056 [cond-mat.str-el]} \BibitemShut {NoStop}%
\bibitem [{\citenamefont {Jackeli}\ and\ \citenamefont
  {Avella}(2015)}]{Jackeli2015}%
  \BibitemOpen
  \bibfield  {author} {\bibinfo {author} {\bibfnamefont {G.}~\bibnamefont
  {Jackeli}}\ and\ \bibinfo {author} {\bibfnamefont {A.}~\bibnamefont
  {Avella}},\ }\href {\doibase 10.1103/PhysRevB.92.184416} {\bibfield
  {journal} {\bibinfo  {journal} {Phys. Rev. B}\ }\textbf {\bibinfo {volume}
  {92}},\ \bibinfo {pages} {184416} (\bibinfo {year} {2015})}\BibitemShut
  {NoStop}%
\bibitem [{\citenamefont {Avella}\ \emph {et~al.}(2018)\citenamefont {Avella},
  \citenamefont {Ciolo},\ and\ \citenamefont {Jackeli}}]{Avella2018}%
  \BibitemOpen
  \bibfield  {author} {\bibinfo {author} {\bibfnamefont {A.}~\bibnamefont
  {Avella}}, \bibinfo {author} {\bibfnamefont {A.~D.}\ \bibnamefont {Ciolo}}, \
  and\ \bibinfo {author} {\bibfnamefont {G.}~\bibnamefont {Jackeli}},\ }\href
  {\doibase https://doi.org/10.1016/j.physb.2017.10.024} {\bibfield  {journal}
  {\bibinfo  {journal} {Physica B: Condensed Matter}\ }\textbf {\bibinfo
  {volume} {536}},\ \bibinfo {pages} {350 } (\bibinfo {year}
  {2018})}\BibitemShut {NoStop}%
\bibitem [{\citenamefont {Rousochatzakis}\ \emph {et~al.}(2016)\citenamefont
  {Rousochatzakis}, \citenamefont {R\"ossler}, \citenamefont {van~den Brink},\
  and\ \citenamefont {Daghofer}}]{Rousochatzakis2016}%
  \BibitemOpen
  \bibfield  {author} {\bibinfo {author} {\bibfnamefont {I.}~\bibnamefont
  {Rousochatzakis}}, \bibinfo {author} {\bibfnamefont {U.~K.}\ \bibnamefont
  {R\"ossler}}, \bibinfo {author} {\bibfnamefont {J.}~\bibnamefont {van~den
  Brink}}, \ and\ \bibinfo {author} {\bibfnamefont {M.}~\bibnamefont
  {Daghofer}},\ }\href {\doibase 10.1103/PhysRevB.93.104417} {\bibfield
  {journal} {\bibinfo  {journal} {Phys. Rev. B}\ }\textbf {\bibinfo {volume}
  {93}},\ \bibinfo {pages} {104417} (\bibinfo {year} {2016})}\BibitemShut
  {NoStop}%
\bibitem [{\citenamefont {{Seabrook}}\ \emph {et~al.}(2019)\citenamefont
  {{Seabrook}}, \citenamefont {{Baez}},\ and\ \citenamefont
  {{Reuther}}}]{Seabrook2019}%
  \BibitemOpen
  \bibfield  {author} {\bibinfo {author} {\bibfnamefont {E.}~\bibnamefont
  {{Seabrook}}}, \bibinfo {author} {\bibfnamefont {M.~L.}\ \bibnamefont
  {{Baez}}}, \ and\ \bibinfo {author} {\bibfnamefont {J.}~\bibnamefont
  {{Reuther}}},\ }\href@noop {} {\bibfield  {journal} {\bibinfo  {journal}
  {arXiv e-prints}\ ,\ \bibinfo {eid} {arXiv:1911.02276}} (\bibinfo {year}
  {2019})},\ \Eprint {http://arxiv.org/abs/1911.02276} {arXiv:1911.02276
  [cond-mat.str-el]} \BibitemShut {NoStop}%
\bibitem [{\citenamefont {Rousochatzakis}\ \emph {et~al.}(2015)\citenamefont
  {Rousochatzakis}, \citenamefont {Reuther}, \citenamefont {Thomale},
  \citenamefont {Rachel},\ and\ \citenamefont {Perkins}}]{Rousochatzakis2015}%
  \BibitemOpen
  \bibfield  {author} {\bibinfo {author} {\bibfnamefont {I.}~\bibnamefont
  {Rousochatzakis}}, \bibinfo {author} {\bibfnamefont {J.}~\bibnamefont
  {Reuther}}, \bibinfo {author} {\bibfnamefont {R.}~\bibnamefont {Thomale}},
  \bibinfo {author} {\bibfnamefont {S.}~\bibnamefont {Rachel}}, \ and\ \bibinfo
  {author} {\bibfnamefont {N.~B.}\ \bibnamefont {Perkins}},\ }\href {\doibase
  10.1103/PhysRevX.5.041035} {\bibfield  {journal} {\bibinfo  {journal} {Phys.
  Rev. X}\ }\textbf {\bibinfo {volume} {5}},\ \bibinfo {pages} {041035}
  (\bibinfo {year} {2015})}\BibitemShut {NoStop}%
\bibitem [{\citenamefont {Auerbach}(1998)}]{Auerbach_book}%
  \BibitemOpen
  \bibfield  {author} {\bibinfo {author} {\bibfnamefont {A.}~\bibnamefont
  {Auerbach}},\ }\href {https://books.google.co.in/books?id=tiQlKzJa6GEC}
  {\emph {\bibinfo {title} {Interacting Electrons and Quantum Magnetism}}},\
  Graduate Texts in Contemporary Physics\ (\bibinfo  {publisher} {Springer New
  York},\ \bibinfo {year} {1998})\BibitemShut {NoStop}%
\bibitem [{\citenamefont {Koga}\ \emph {et~al.}(2018)\citenamefont {Koga},
  \citenamefont {Tomishige},\ and\ \citenamefont {Nasu}}]{Koga2018}%
  \BibitemOpen
  \bibfield  {author} {\bibinfo {author} {\bibfnamefont {A.}~\bibnamefont
  {Koga}}, \bibinfo {author} {\bibfnamefont {H.}~\bibnamefont {Tomishige}}, \
  and\ \bibinfo {author} {\bibfnamefont {J.}~\bibnamefont {Nasu}},\ }\href
  {\doibase 10.7566/JPSJ.87.063703} {\bibfield  {journal} {\bibinfo  {journal}
  {Journal of the Physical Society of Japan}\ }\textbf {\bibinfo {volume}
  {87}},\ \bibinfo {pages} {063703} (\bibinfo {year} {2018})},\ \Eprint
  {http://arxiv.org/abs/https://doi.org/10.7566/JPSJ.87.063703}
  {https://doi.org/10.7566/JPSJ.87.063703} \BibitemShut {NoStop}%
\bibitem [{\citenamefont {Oitmaa}\ \emph {et~al.}(2018)\citenamefont {Oitmaa},
  \citenamefont {Koga},\ and\ \citenamefont {Singh}}]{Oitmaa2018}%
  \BibitemOpen
  \bibfield  {author} {\bibinfo {author} {\bibfnamefont {J.}~\bibnamefont
  {Oitmaa}}, \bibinfo {author} {\bibfnamefont {A.}~\bibnamefont {Koga}}, \ and\
  \bibinfo {author} {\bibfnamefont {R.~R.~P.}\ \bibnamefont {Singh}},\ }\href
  {\doibase 10.1103/PhysRevB.98.214404} {\bibfield  {journal} {\bibinfo
  {journal} {Phys. Rev. B}\ }\textbf {\bibinfo {volume} {98}},\ \bibinfo
  {pages} {214404} (\bibinfo {year} {2018})}\BibitemShut {NoStop}%
\bibitem [{\citenamefont {Suzuki}\ and\ \citenamefont
  {Yamaji}(2018)}]{Suzuki2018}%
  \BibitemOpen
  \bibfield  {author} {\bibinfo {author} {\bibfnamefont {T.}~\bibnamefont
  {Suzuki}}\ and\ \bibinfo {author} {\bibfnamefont {Y.}~\bibnamefont
  {Yamaji}},\ }\href {\doibase https://doi.org/10.1016/j.physb.2017.09.105}
  {\bibfield  {journal} {\bibinfo  {journal} {Physica B: Condensed Matter}\
  }\textbf {\bibinfo {volume} {536}},\ \bibinfo {pages} {637 } (\bibinfo {year}
  {2018})}\BibitemShut {NoStop}%
\bibitem [{\citenamefont {Minakawa}\ \emph {et~al.}(2019)\citenamefont
  {Minakawa}, \citenamefont {Nasu},\ and\ \citenamefont {Koga}}]{Minakawa2019}%
  \BibitemOpen
  \bibfield  {author} {\bibinfo {author} {\bibfnamefont {T.}~\bibnamefont
  {Minakawa}}, \bibinfo {author} {\bibfnamefont {J.}~\bibnamefont {Nasu}}, \
  and\ \bibinfo {author} {\bibfnamefont {A.}~\bibnamefont {Koga}},\ }\href
  {\doibase 10.1103/PhysRevB.99.104408} {\bibfield  {journal} {\bibinfo
  {journal} {Phys. Rev. B}\ }\textbf {\bibinfo {volume} {99}},\ \bibinfo
  {pages} {104408} (\bibinfo {year} {2019})}\BibitemShut {NoStop}%
\bibitem [{\citenamefont {Stavropoulos}\ \emph {et~al.}(2019)\citenamefont
  {Stavropoulos}, \citenamefont {Pereira},\ and\ \citenamefont
  {Kee}}]{Stavropoulos2019}%
  \BibitemOpen
  \bibfield  {author} {\bibinfo {author} {\bibfnamefont {P.~P.}\ \bibnamefont
  {Stavropoulos}}, \bibinfo {author} {\bibfnamefont {D.}~\bibnamefont
  {Pereira}}, \ and\ \bibinfo {author} {\bibfnamefont {H.-Y.}\ \bibnamefont
  {Kee}},\ }\href {\doibase 10.1103/PhysRevLett.123.037203} {\bibfield
  {journal} {\bibinfo  {journal} {Phys. Rev. Lett.}\ }\textbf {\bibinfo
  {volume} {123}},\ \bibinfo {pages} {037203} (\bibinfo {year}
  {2019})}\BibitemShut {NoStop}%
\bibitem [{\citenamefont {Chaloupka}\ \emph {et~al.}(2010)\citenamefont
  {Chaloupka}, \citenamefont {Jackeli},\ and\ \citenamefont
  {Khaliullin}}]{Chaloupka2010}%
  \BibitemOpen
  \bibfield  {author} {\bibinfo {author} {\bibfnamefont {J.}~\bibnamefont
  {Chaloupka}}, \bibinfo {author} {\bibfnamefont {G.}~\bibnamefont {Jackeli}},
  \ and\ \bibinfo {author} {\bibfnamefont {G.}~\bibnamefont {Khaliullin}},\
  }\href {\doibase 10.1103/PhysRevLett.105.027204} {\bibfield  {journal}
  {\bibinfo  {journal} {Phys. Rev. Lett.}\ }\textbf {\bibinfo {volume} {105}},\
  \bibinfo {pages} {027204} (\bibinfo {year} {2010})}\BibitemShut {NoStop}%
\bibitem [{\citenamefont {Price}\ and\ \citenamefont
  {Perkins}(2013)}]{Price2013}%
  \BibitemOpen
  \bibfield  {author} {\bibinfo {author} {\bibfnamefont {C.}~\bibnamefont
  {Price}}\ and\ \bibinfo {author} {\bibfnamefont {N.~B.}\ \bibnamefont
  {Perkins}},\ }\href {\doibase 10.1103/PhysRevB.88.024410} {\bibfield
  {journal} {\bibinfo  {journal} {Phys. Rev. B}\ }\textbf {\bibinfo {volume}
  {88}},\ \bibinfo {pages} {024410} (\bibinfo {year} {2013})}\BibitemShut
  {NoStop}%
\bibitem [{\citenamefont {Khatua}\ \emph {et~al.}(2018)\citenamefont {Khatua},
  \citenamefont {Shankar},\ and\ \citenamefont {Ganesh}}]{Khatua2018}%
  \BibitemOpen
  \bibfield  {author} {\bibinfo {author} {\bibfnamefont {S.}~\bibnamefont
  {Khatua}}, \bibinfo {author} {\bibfnamefont {R.}~\bibnamefont {Shankar}}, \
  and\ \bibinfo {author} {\bibfnamefont {R.}~\bibnamefont {Ganesh}},\ }\href
  {\doibase 10.1103/PhysRevB.97.054403} {\bibfield  {journal} {\bibinfo
  {journal} {Phys. Rev. B}\ }\textbf {\bibinfo {volume} {97}},\ \bibinfo
  {pages} {054403} (\bibinfo {year} {2018})}\BibitemShut {NoStop}%
\bibitem [{\citenamefont {Pauling}(1936)}]{Pauling1936}%
  \BibitemOpen
  \bibfield  {author} {\bibinfo {author} {\bibfnamefont {L.}~\bibnamefont
  {Pauling}},\ }\href {\doibase 10.1063/1.1749766} {\bibfield  {journal}
  {\bibinfo  {journal} {The Journal of Chemical Physics}\ }\textbf {\bibinfo
  {volume} {4}},\ \bibinfo {pages} {673} (\bibinfo {year} {1936})}\BibitemShut
  {NoStop}%
\bibitem [{\citenamefont {Kottos}\ and\ \citenamefont
  {Smilansky}(1997)}]{Kottos1997}%
  \BibitemOpen
  \bibfield  {author} {\bibinfo {author} {\bibfnamefont {T.}~\bibnamefont
  {Kottos}}\ and\ \bibinfo {author} {\bibfnamefont {U.}~\bibnamefont
  {Smilansky}},\ }\href {\doibase 10.1103/PhysRevLett.79.4794} {\bibfield
  {journal} {\bibinfo  {journal} {Phys. Rev. Lett.}\ }\textbf {\bibinfo
  {volume} {79}},\ \bibinfo {pages} {4794} (\bibinfo {year}
  {1997})}\BibitemShut {NoStop}%
\bibitem [{\citenamefont {Keating}(2008)}]{Keating2008}%
  \BibitemOpen
  \bibfield  {author} {\bibinfo {author} {\bibfnamefont {J.}~\bibnamefont
  {Keating}},\ }in\ \href {https://books.google.co.in/books?id=k04FCAAAQBAJ}
  {\emph {\bibinfo {booktitle} {Analysis on Graphs and Its Applications: Isaac
  Newton Institute for Mathematical Sciences, Cambridge, UK, January 8-June 29,
  2007}}},\ \bibinfo {series and number} {Proceedings of symposia in pure
  mathematics},\ \bibinfo {editor} {edited by\ \bibinfo {editor} {\bibfnamefont
  {P.}~\bibnamefont {Exner}}, \bibinfo {editor} {\bibfnamefont
  {J.}~\bibnamefont {Keating}}, \bibinfo {editor} {\bibfnamefont
  {P.}~\bibnamefont {Kuchment}}, \bibinfo {editor} {\bibfnamefont
  {A.}~\bibnamefont {Teplyaev}}, \ and\ \bibinfo {editor} {\bibfnamefont
  {T.}~\bibnamefont {Sunada}}}\ (\bibinfo  {publisher} {American Mathematical
  Society},\ \bibinfo {year} {2008})\ pp.\ \bibinfo {pages}
  {291--314}\BibitemShut {NoStop}%
\bibitem [{\citenamefont {Harrison}\ \emph {et~al.}(2011)\citenamefont
  {Harrison}, \citenamefont {Keating},\ and\ \citenamefont
  {Robbins}}]{Harrison2011}%
  \BibitemOpen
  \bibfield  {author} {\bibinfo {author} {\bibfnamefont {J.~M.}\ \bibnamefont
  {Harrison}}, \bibinfo {author} {\bibfnamefont {J.~P.}\ \bibnamefont
  {Keating}}, \ and\ \bibinfo {author} {\bibfnamefont {J.~M.}\ \bibnamefont
  {Robbins}},\ }\href {\doibase 10.1098/rspa.2010.0254} {\bibfield  {journal}
  {\bibinfo  {journal} {Proceedings of the Royal Society A: Mathematical,
  Physical and Engineering Sciences}\ }\textbf {\bibinfo {volume} {467}},\
  \bibinfo {pages} {212} (\bibinfo {year} {2011})}\BibitemShut {NoStop}%
\bibitem [{\citenamefont {Alexandradinata}\ and\ \citenamefont
  {Glazman}(2018)}]{Alexandradinata2018}%
  \BibitemOpen
  \bibfield  {author} {\bibinfo {author} {\bibfnamefont {A.}~\bibnamefont
  {Alexandradinata}}\ and\ \bibinfo {author} {\bibfnamefont {L.}~\bibnamefont
  {Glazman}},\ }\href {\doibase 10.1103/PhysRevB.97.144422} {\bibfield
  {journal} {\bibinfo  {journal} {Phys. Rev. B}\ }\textbf {\bibinfo {volume}
  {97}},\ \bibinfo {pages} {144422} (\bibinfo {year} {2018})}\BibitemShut
  {NoStop}%
\end{thebibliography}%
\end{document}